\documentstyle[psfig,macros]{mn}
\begin{document}
%

\title[Statistics of Dark Matter Substructure I]
      {Statistics of Dark Matter Substructure: I.
       Model and Universal Fitting Functions}

\author[Jiang \& van den Bosch]
       {Fangzhou Jiang\thanks{E-mail:fangzhou.jiang@yale.edu} 
        \& Frank C. van den Bosch \\
        Department of Astronomy, Yale University, New Haven, CT 06511, USA}


\date{}

\pagerange{\pageref{firstpage}--\pageref{lastpage}}
\pubyear{2013}

\maketitle

\label{firstpage}


\begin{abstract}
We present a new, semi-analytical model describing the evolution of dark matter subhaloes. The model uses merger trees constructed using the method of Parkinson et al. (2008) to describe the masses and redshifts of subhaloes at accretion, which are subsequently evolved using a simple model for the orbit-averaged mass loss rates. The model is extremely fast, treats subhaloes of all orders, accounts for scatter in orbital properties and halo concentrations, and uses a simple recipe to convert subhalo mass to maximum circular velocity. The model accurately reproduces the average subhalo mass and velocity functions in numerical simulations. The inferred subhalo mass loss rates imply that an average dark matter subhalo loses in excess of 80 percent of its infall mass during its first radial orbit within the host halo. We demonstrate that the total mass fraction in subhaloes is tightly correlated with the `dynamical age' of the host halo, defined as the number of halo dynamical times that have elapsed since its formation. Using this relation, we present universal fitting functions for the evolved and unevolved subhalo mass and velocity functions that are valid for any host halo mass, at any redshift, and for any $\Lambda$CDM cosmology.
\end{abstract} 




\section{Introduction} 
\label{Sec:Introduction}

Numerical $N$-body simulations have shown that when two dark matter
haloes merge, the less massive progenitor halo initially survives as a
self-bound entity, called a subhalo, orbiting within the potential
well of the more massive progenitor halo. These subhaloes are
subjected to tidal forces and impulsive encounters with other
subhaloes causing tidal heating and mass stripping, and to dynamical
friction that causes them to lose orbital energy and angular momentum
to the dark matter particles of the `host' halo.  Depending on its
orbit, density profile, and mass, a subhalo therefore either survives
to the present day or is disrupted; the operational distinction being
whether a self-bound entity remains or not.

Characterizing the statistics and properties of dark matter subhaloes
is of paramount importance for various areas of astrophysics. First of
all, subhaloes are believed to host satellite galaxies, and the
abundance of satellite galaxies is therefore directly related to that
of subhaloes. This basic idea underlies the popular technique of
subhalo abundance matching (e.g., Vale \& Ostriker 2004; Conroy \etal
2006, 2007; Guo \etal 2011; Hearin \etal 2013) and has given rise to
two problems in our understanding of galaxy formation: the ``missing
satellite" problem (Moore \etal 1999; Klypin \etal 1999) and the ``too
big to fail" problem (Boylan-Kolchin \etal 2011). Secondly,
substructure is also important in the field of gravitational lensing,
where it can cause time-delays (e.g., Keeton \& Moustakas 2009) and
flux-ratio anomalies (Metcalf \& Madau 2001; Brada\v{c} \etal 2002;
Dalal \& Kochanek 2002), and for the detectability of dark matter
annihilation, where the clumpiness due to substructure is responsible
for a `boost factor' (e.g., Diemand \etal 2007; Pieri \etal 2008;
Giocoli \etal 2008b). Finally, the abundance and properties of dark
matter substructure controls the survivability of fragile structures
in dark matter haloes, such as tidal streams and/or galactic disks
(T\'oth \& Ostriker 1992; Taylor \& Babul 2001; Ibata \etal 2002;
Carlberg 2009).

The most common statistic used to describe the substructure of dark
matter haloes is the subhalo mass function (hereafter SHMF), $\rmd
N/\rmd\ln(m/M)$, which expresses the (average) number of subhaloes of
mass $m$ per host halo of mass $M$, per logarithmic bin of
$m/M$. Following van den Bosch, Tormen \& Giocoli (2005), we will
distinguish two different SHMFs; the {\it unevolved} SHMF, where $m$
is the mass of the subhalo {\it at accretion}, and the {\it evolved}
SHMF, where $m$ reflects the mass of the surviving, self-bound entity
at the present day, which is reduced with respect to that at accretion
due to mass stripping.

The SHMFs of dark matter haloes have been studied using two
complementary techniques; $N$-body simulations (e.g., Tormen 1997;
Tormen, Diaferio \& Syer 1998; Moore \etal 1998, 1999; Klypin \etal
1999a,b; Ghigna \etal 1998, 2000; Stoehr \etal 2002; De Lucia \etal
2004; Diemand, Moore \& Stadel 2004; Gill \etal 2004a,b; Gao \etal
2004; Reed \etal 2005; Kravtsov \etal 2004; Giocoli \etal 2008a, 2010;
Weinberg \etal 2008) and semi-analytical techniques based on the
extended Press-Schechter (EPS; Bond \etal 1991) formalism (e.g.,
Taylor \& Babul 2001, 2004, 2005a,b; Benson \etal 2002; Taffoni \etal
2003; Oguri \& Lee 2004; Zentner \& Bullock 2003; Pe\~{n}arrubia \&
Benson 2005; Zentner \etal 2005; van den Bosch \etal 2005; Gan \etal
2010; Yang \etal 2011; Purcell \& Zentner 2012).  Both methods have
their own pros and cons. Numerical simulations have the virtue of
including all relevant, gravitational physics related to the assembly
of dark matter haloes, and the evolution of the subhalo
population. However, they are also extremely CPU intensive, and the
results depend on the mass- and force-resolution adopted.  In
addition, there is some level of arbitrariness in how to identify
haloes and subhaloes in the simulations. In particular, different
(sub)halo finders applied to the same simulation output typically
yield subhalo mass functions that differ at the 10-20 percent level
(Knebe \etal 2011, 2013; Onions \etal 2012) or more (van den Bosch \&
Jiang 2014). Semi-analytical techniques, on the other hand, don't
suffer from issues related to subhalo identification or force
resolution, and are significantly faster, but their downside is that
the relevant physics is only treated approximately.

All semi-analytical methods require two separate ingredients: halo merger
trees, which describe the hierarchical assembly of dark matter haloes,
and a treatment of the various physical processes that cause the
subhalo population to evolve (dynamical friction, tidal heating and
stripping, impulsive encounters).  To properly account for all these
processes, which depend strongly on the orbital properties, requires a
detailed integration over all individual subhalo orbits. This is
complicated by the fact that the mass of the parent halo evolves with
time.  If the mass growth rate is sufficiently slow, the evolution may
be considered adiabatic, thus allowing the orbits of subhaloes to be
integrated analytically despite the non-static nature of the
background potential. This principle is exploited in many of the
semi-analytical based models listed above.  In reality, however,
haloes grow hierarchically through (major) mergers, making the actual
orbital evolution highly non-linear.

In order to sidestep these difficulties, van den Bosch, Tormen \&
Giocoli (2005; hereafter B05) considered the {\it average} mass loss
rate of dark matter subhaloes, where the average is taken over the
entire distribution of orbital configurations (energies, angular
momenta, and orbital phases). This removes the requirement to actually
integrate individual orbits, allowing for an extremely fast
calculation of the evolved subhalo mass function. B05 adopted a simple
functional form for the average mass loss rate, which had two free
parameters which they calibrated by comparing the resulting subhalo
mass functions to those obtained using numerical simulations. In a
subsequent paper, Giocoli, Tormen \& van den Bosch (2008; hereafter
G08), directly measured the average mass loss rate of dark matter
subhaloes in numerical simulations.  They found that the functional
form adopted by B05 adequately describes the average mass loss rates
in the simulations, but with best-fit values for the free parameters
that are substantially different. G08 argued that this discrepancy
arises from the fact that B05 used the `N-branch method with
accretion' of Somerville \& Kolatt (1999; hereafter SK99) to construct
their halo merger trees, which results in an unevolved subhalo mass
function that is significantly different from what is found in the
simulations. This was recently confirmed by the authors in a detailed
comparison of merger tree algorithms (Jiang \& van den Bosch 2014a;
hereafter JB14). Note that this same SK99 method has also been used by
most of the other semi-analytical models for dark matter substructure,
including Taylor \& Babul (2004, 2005a,b), Zentner \& Bullock (2003),
Zentner \etal (2005) and even the recent study by Purcell \& Zentner
(2012).

In this series of papers we use an overhauled version of the
semi-analytical method pioneered by B05 to study the statistics of dark
matter subhaloes in unprecedented detail.  In particular, we extent
and improve upon B05 by (i) using halo merger trees constructed with
the more reliable method of Parkinson, Cole \& Helly (2008), (ii)
evolving subhaloes using the improved mass-loss model of G08 and
accounting for stochasticity in the mass-loss rates due to the scatter
in orbital properties and halo concentrations, (iii) considering the
entire hierarchy of dark matter subhaloes (including sub-subhaloes,
sub-sub-subhaloes, etc.), and (iv) predicting not only the masses of
subhaloes but also their maximum circular velocities, $V_{\rm max}$.
In this paper, the first in the series, we present the improved
semi-analytical model, followed by a detailed study of the (average)
subhalo abundance as function of mass and maximum circular velocity
including a presentation of universal fitting functions. In Paper II
(van den Bosch \& Jiang 2014) we present a more detailed comparison of
the model predictions with simulation results, paying special
attention to the large discrepancies among different simulation
results that arise from the use of different subhalo finders. Finally,
in Paper III (Jiang \& van den Bosch; in preparation) we exploit our
semi-analytical model to quantify the halo-to-halo variation of populations
of dark matter subhaloes.

This paper is organized as follows. We start in \S\ref{Sec:Model} with
a detailed description of our semi-analytical model, including the
construction of halo merger trees (\S\ref{Sec:Trees}), an updated
model for the average mass loss rate of subhaloes
(\S\ref{Sec:MassLoss}), and a description of how we convert (sub)halo
masses to their corresponding $V_{\rm max}$ (\S\ref{Sec:Vmax}). In
\S\ref{Sec:Test} we demonstrate that the model can accurately
reproduce both the subhalo mass and velocity functions obtained from
numerical simulations, after tuning our single free model parameter,
and we discuss the scalings with host halo mass and
redshift. \S\ref{Sec:Universal} presents accurate, universal fitting
functions for the average subhalo mass and velocity functions that are
valid for any host halo mass, redshift and $\Lambda$CDM cosmology. In
\S\ref{Sec:Discussion} we discuss implications of our inferred subhalo
mass loss rates, and we summarize our results in \S\ref{Sec:Summary}.

Throughout we use $m$ and $M$ to refer to the masses of subhaloes and
host haloes, respectively, use $\ln$ and $\log$ to indicate the natural
logarithm and 10-based logarithm, respectively, and express units that
depend on the Hubble constant in terms of $h = H_0/(100\kmsmpc)$.


\section{Model Description}  
\label{Sec:Model}

The goal of this series of papers is to present a study of
unprecedented detail regarding the statistics of dark matter
subhaloes. To that extent we use an improved and extended version of
the semi-analytical model of B05 which computes the masses of subhaloes 
for a particular realization of a host halo's mass assembly history. In
what follows we present a detailed description of the model, including
a discussion of how and where we make improvements, and add
extensions, with respect to the B05 model.

\subsection{Halo Merger Trees}
\label{Sec:Trees}

The backbone of the B05 model, and of any other semi-analytical model
for the substructure of dark matter haloes, is halo merger trees.
These describe the hierarchical mass assembly of dark matter haloes,
and therefore yield the masses and redshifts at which dark matter
subhaloes are accreted into their hosts.
\begin{figure}
\centerline{\psfig{figure=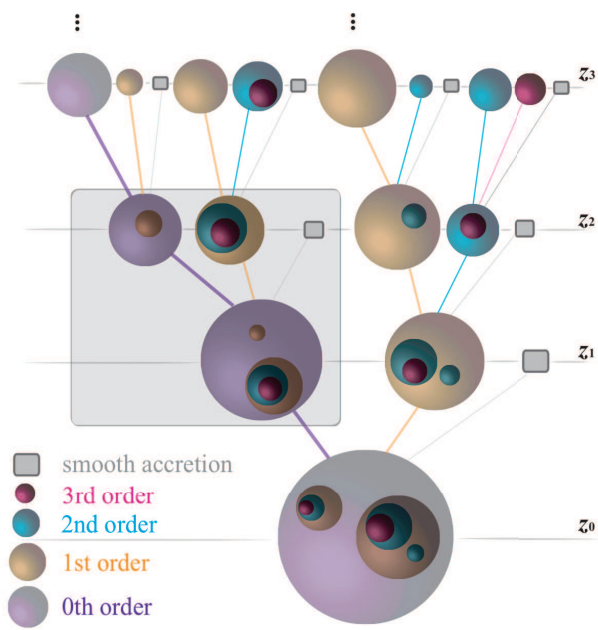,width=\hssize}}
\caption{Illustration depicting the anatomy of a merger tree for a
  host halo (purple sphere at the bottom) at redshift $z=z_0$. The
  purple spheres to the left illustrate the assembly history of the
  main progenitor. We refer to these as `zeroth-order' progenitors,
  and they accrete `first-order' progenitors, which end up as
  (first-order) subhaloes at $z=z_0$. In turn, these first-order
  progenitors accrete second-order progenitors which end-up as
  second-order subhaloes (sub-subhaloes) at $z=z_0$, etc.  The size of
  a sphere is proportional to its mass, while its color reflects its
  order, as indicated.  The large, shaded box highlights a single
  branching point in the tree structure, which shows a descendant halo
  plus its single-time-step progenitors.  The small shaded boxes
  present at each branching point reflect `smooth accretion', as
  defined in the text.}
\label{fig:mergertree}
\end{figure}

Before describing the construction of our merger trees, it is useful to
introduce some terminology that is used throughout this paper.
Fig.~\ref{fig:mergertree} shows a schematic representation of a merger
tree. We refer to the halo at the base of the tree (i.e., the large
purple halo at $z=z_0$) as the {\it host halo}. For each individual
branching point along the tree (one example is highlighted in
Fig.~\ref{fig:mergertree}), the end-product of the merger event is
called the {\it descendant halo}, while the haloes that merge are
called the {\it progenitors}. The {\it main progenitor} of a
descendant halo is the progenitor that contributes the most mass. For
example, for the branching point highlighted in
Fig.~\ref{fig:mergertree}, the purple halo at $z=z_2$ is the main
progenitor of its descendant at $z=z_1$.  The {\it main branch} of the
merger tree is defined as the branch tracing the main progenitor of
the main progenitor of the main progenitor, etc. (i.e., the branch
connecting the purple haloes). Note that the main progenitor halo at
redshift $z$ is not necessarily the most massive progenitor at that
redshift. Throughout we shall occasionally refer to the main
progenitor haloes of a given host halo as its zeroth-order
progenitors, while the mass history, $M(z)$, along this branch is
called the {\it mass assembly history} (MAH). Haloes that accrete
directly onto the main branch are called first-order progenitors, or,
after accretion, first-order subhaloes. Similarly, haloes that accrete
directly onto first-order progenitors are called second-order
progenitors, and they end up at $z=z_0$ as second-order subhaloes (or
sub-subhaloes) of the host halo. The same logic is used to define
higher-order progenitors and subhaloes, as illustrated in
Fig.~\ref{fig:mergertree}. An $n^{\rm th}$-order (sub)halo that hosts
an $(n+1)^{\rm th}$-order subhalo is called a {\it parent} halo of the
$(n+1)^{\rm th}$-order subhalo. Throughout, we define (sub)halo masses
such that the mass of an $n^{\rm th}$-order parent halo {\it includes}
the masses of its subhaloes of order $n+1$; we refer to this as the
{\it inclusive} definition of subhalo mass. Most subhalo finders
used to analyze $N$-body simulations use the same inclusive definition of
subhalo mass, though not all of them (see Paper~II for a detailed
discussion).

We construct our merger trees using the EPS formalism, which has the
advantage over using numerical simulations that it is not hampered by
ambiguities having to do with the identification and linking of
(sub)haloes. EPS provides the progenitor mass function (hereafter
PMF), $n_{\rm EPS}(M_\rmp, z_1|M_0,z_0)$, describing the
ensemble-average number, $n_{\rm EPS}(M_\rmp, z_1|M_0,z_0)\rmd
M_\rmp$, of progenitors of mass $M_\rmp$ that a descendant halo of
mass $M_0$ at redshift $z_0$ has at redshift $z_1 > z_0$. Starting
from some target host halo mass $M_0$ at $z_0$, one can use this PMF
to draw a set of progenitor masses ${M_{\rmp,1},
  M_{\rmp,2},...,M_{\rmp,N}}$ at some earlier time $z_1 = z_0 + \Delta
z$, where $\sum_{i=1}^{N} M_{\rmp,i} = M_0$ in order to assure mass
conservation. The time-step $\Delta z$ used sets the `temporal
resolution' of the merger tree, and may vary along the tree. This
procedure is then repeated for each progenitor with mass $M_{\rmp,i} >
M_{\rm res}$, thus advancing `upwards' along the tree. The minimum
mass $M_{\rm res}$ sets the `mass resolution' of the merger tree and
is typically expressed as a fraction of the final host mass $M_0$.
The small shaded boxes in Fig.~\ref{fig:mergertree}, present at each
branching, reflect the mass accreted by the descendant halo in the
form of smooth accretion (i.e., not part of any halo) or in the form
of progenitor haloes with masses $M_\rmp < M_{\rm res}$. Throughout we
shall refer to this component as {\it smooth accretion}.

Haloes at the top of the tree that have no progenitors with mass
$M_{\rmp} > M_{\rm res}$ are called the {\it leave} haloes of the
tree, and the mass evolution of a leave halo down to $z=0$ is called
the halo's {\it trajectory}. Only one trajectory per tree corresponds
to a host halo (namely that of the main progenitor), while all other
trajectories end up as sub-haloes at $z=0$ (of different orders).  The
moment a halo is for the first time accreted into a more massive halo
(i.e., transits from being a host halo to a sub-halo) is called the
halo's {\it accretion time}, $t_{\rm acc}$, and it's mass at that time is
called the {\it accretion mass}, for which we use $m_{\rm acc}$ or $m_{\rm
  acc}$ without distinction.  Throughout we use the subscript `0' to
refer to properties at redshift $z=z_0$ (typically we adopt $z_0=0$);
Hence, $m_0$ is the $z=z_0$ mass of a sub-halo, which differs from
$m_{\rm acc}$ due to mass loss (see \S\ref{Sec:MassLoss} below).  In what
follows we refer to the mass functions $\rmd N/\rmd\ln(m_0/M_0)$ and
$\rmd N / \rmd\ln(m_{\rm acc}/M_0)$ as the {\it evolved} and {\it
  unevolved} SHMFs of a host halo of mass $M_0$, respectively.

B05 constructed EPS merger trees using the SK99 method, but only up to
first order; i.e., they only considered first-order subhaloes, and
therefore did not require merger trees that resolve the assembly
histories of first-order progenitors. We extent this by using full
merger trees, thus allowing us to study the statistics of subhaloes of
all orders. In addition, we also improve upon B05 by using another
method to construct our merger trees. In JB14 we tested and compared a
number of different Monte Carlo algorithms for constructing EPS-based
merger trees, including SK99. We showed that the SK99-method results
in (i) haloes that assemble too late, (ii) merger rates that are too
high by factors of two to three (see also Fakhouri \& Ma 2008; Genel
\etal 2009), and (iii) unevolved subhalo mass functions that are much
too high, especially for subhaloes with $m_{\rm acc} \sim M_0/100$. As
first pointed out by G08, and as discussed in more detail in JB14,
this explains why B05 inferred an average subhalo mass loss rate that
is too high. It also implies that other models for dark matter
substructure that are based on the SK99 algorithm (Taylor \& Babul
2004, 2005; Zentner \& Bullock 2003; Zentner \etal 2005; Purcell \&
Zentner 2012), are likely to suffer from similar systematic errors.

As we demonstrated in JB14, the merger tree algorithm developed by
Cole \etal (2000), which is used in the semi-analytical substructure
models of Benson \etal (2002) and Pe\~{n}arrubia \& Benson (2005), has
similar shortcomings as SK99, albeit at a significantly reduced
level. However, it still overpredicts the unevolved SHMF by $\sim 40$
percent, and is therefore not well suited to model the population of
dark matter subhaloes. Of all the methods tested by JB14, the one that
clearly stood out as the most reliable is that of Parkinson \etal
(2008; hereafter P08). The P08 algorithm is a modification of the
binary algorithm of Cole \etal (2000), in which the PMF is modified
with respect to the EPS prediction to match results from the Millennium
simulation (see Cole \etal 2008). As shown in JB14, the P08 algorithm
yields merger rates and unevolved SHMFs that are in excellent
agreement with simulation results within the errors.  Hence, in this
paper we improve upon B05 by using merger trees constructed with the
P08 algorithm. Throughout we always adopt a mass resolution of
$\psi_{\rm res} \equiv M_{\rm res}/M_0 = 10^{-5}$ unless mentioned
otherwise, and construct the merger trees using the time stepping
advocated in Appendix~A of P08 (which roughly corresponds to $\Delta
z\sim10^{-3}$; somewhat finer/coarser at high/low redshift). In order
to speed up the code, and to reduce memory requirements, we down-sample
the time resolution of each merger tree by registering progenitor
haloes every time step $\Delta t = 0.1 t_{\rm ff}(z)$. Here $t_{\rm
  ff}(z) \propto (1+z)^{-3/2}$ is the free-fall time for a halo with
an overdensity of 200 at redshift $z$. Since the orbital time of a
subhalo is of order the free-fall time, there is little added value in
resolving merger trees at higher time resolution than this. We have
verified that indeed our results do not change if we use merger trees
with a smaller time step.

\subsection{Subhalo Mass Evolution}
\label{Sec:MassLoss}

The next ingredient in the semi-analytical method is a model for the mass
evolution of the subhalo as it orbits its host halo. This is governed
by tidal stripping, tidal heating, dynamical friction, and the
impulsive heating due to high-speed encounters with other
substructures. Consequently, the mass evolution of a subhalo can vary
dramatically along an orbit, and also depends strongly on the orbital
energy and angular momentum (see e.g., Taffoni \etal 2003; Taylor \&
Babul 2004; Penarrubia \& Benson 2005; Zentner \etal 2005; Gan \etal
2010).

The unique aspect of the B05 approach is that it considers the {\it
  average} mass loss rate of a dark matter subhalo, where the average
is taken over all orbital energies, eccentricities and phases. Using
the fact that dark matter haloes have a universal density profile
(e.g., Navarro, Frenk \& White 1997), and that the distribution of
orbital properties of infalling subhaloes have only a mild dependence
on parent halo mass (e.g, Zentner \etal 2005; Wang \etal 2005;
Khochfar \& Burkert 2006; Wetzel 2011), this average mass loss rate,
to first order, only depends on the instantaneous masses of the
subhalo, $m$, and parent halo, $M$. In fact, B05 postulated that the
dependence on parent halo mass only enters through the (instantaneous)
mass ratio $m/M$, i.e.,
\begin{equation}\label{mydecay}
\dot{m} = - \calA \, {m \over \tau_{\rm dyn}} \left({m\over M}\right)^{\zeta}.
\end{equation}
Here the negative sign is to emphasize that $m$ is expected to decrease with
time, $\calA$ and $\zeta$ are two free parameters describing the normalization
and mass dependence of the subhalo mass loss rate, respectively, and 
\begin{eqnarray}\label{mytau}
\tau_{\rm dyn}(z) & = & \sqrt{3\, \pi \over 16\,G\,\bar{\rho}_\rmh(z)}\nonumber \\
& = & 1.628 h^{-1} {\rm Gyr} \,
\left[{\Delta_{\rm vir}(z) \over 178}\right]^{-1/2} \, 
\left[ {H(z) \over H_0} \right]^{-1} 
\end{eqnarray}
is the halo's dynamical time, with $H(z)$ the Hubble parameter, and
$\Delta_{\rm vir}(z)$ the virial parameter that expresses the average
density of a virialized dark matter halo, $\bar{\rho}_\rmh$, at
redshift $z$ in units of the critical density at that
redshift. Throughout this paper we adopt the fitting function for
$\Delta_{\rm vir}(z)$ given by Bryan \& Norman (1998).

For a subhalo embedded in a static parent halo ($\dot{M}=0$) this
yields
\begin{equation}
\label{myevolv}
m(t+\Delta t) = \left\{ \begin{array}{ll}
m(t) \, \exp(-\Delta t/\tau) & \mbox{if $\zeta = 0$} \\
m(t) \left[ 1 + \zeta \, \left( {m\over M}\right)^\zeta \, 
\left({\Delta t\over \tau}\right) \right]^{-1/\zeta} & \mbox{otherwise} 
\end{array} \right.
\end{equation}
where $\tau = \tau(z) \equiv \tau_{\rm dyn}(z)/\calA$ is the
characteristic mass-loss time scale at redshift $z$.

Although the subhalo mass loss rate in a static halo is a well defined
concept, in reality the parent mass $M$ increases with time due to the
accretion of matter and other subhaloes.  Following B05, we utilize
the discrete time stepping of the merger tree to evolve both $m$ and
$M$. At the beginning of each time step the parent halo is assumed to
instantaneously increase its mass as described by the merger tree
($\dot{M} > 0, \dot{m}=0$), while in between two time steps we set
$\dot{M}=0$ and evolve the subhalo mass, $m$, according to
Eq.~(\ref{myevolv}).  In what follows we consider it understood that
$m$ and $M$ depend on time, without having to write this
time-dependence explicitly.

B05 adopted the subhalo mass loss model given by Eqs.~(\ref{mydecay})
and (\ref{mytau}) and tuned the free parameters, $\calA$ and $\zeta$,
such that their evolved SHMF matched that obtained from simulations.
This resulted in $\calA = 23.7$, corresponding to a characteristic
mass-loss time scale at $z=0$ equal to $\tau_0 = 0.13$Gyr, and $\zeta
= 0.36$. Using high-resolution $N$-body simulations, G08 computed the
{\it average} mass loss rates of dark matter subhaloes as a function
of both halo mass and redshift.  They found that the functional forms
of Eqs.~(\ref{mydecay}) and (\ref{mytau}) adequately describes the
simulation results, but with best-fit values $\calA =
1.54^{+0.52}_{-0.31}$ (corresponding to $\tau_0 = (2.0 \pm 0.5)$Gyr)
and $\zeta = 0.07 \pm 0.03$ that are very different from those
obtained in B05. In particular, the much smaller value for $\calA$
implies significantly lower mass loss rates. As discussed in G08, this
is a manifestation of the shortcomings of the SK99 algorithm used by
B05 to construct their merger trees; in order to match the evolved
SHMF in the simulations, B05 had to adopt higher mass loss rates to
compensate for the overabundance of accreted subhaloes.

In this paper we adopt $\zeta = 0.07$ and treat $\calA$ as a free
parameter, which we tune by fitting our evolved subhalo mass functions
to simulation results. As we show, mainly because we now use more
accurate halo merger trees, the resulting value of $\calA$ ($1.34$)
is in excellent agreement with the simulation results of G08.

\subsubsection{Subhalo Mass Loss Rates; a toy model}
\label{Sec:ToyModel}

An important goal of this work (discussed in detail in Paper~II) is to
assess the halo-to-halo variance of subhalo statistics. The main
source of this variance is the scatter in mass accretion histories
(i.e., halo-to-halo variance in the accretion masses and accretion
redshifts of subhaloes), which is accounted for via our merger
trees. However, another important source of scatter is that due to
variance in the orbital properties; orbits with more negative orbital
energy, or with smaller angular momentum, will have a smaller
pericenter and therefore experience more tidal stripping.  In
addition, since the tidal radius of a subhalo depends on the density
profiles of host halo and subhalo, a third source of scatter is the
variance in halo concentrations, of both the host halo and the subhalo.
Using numerical simulations, G08 indeed found a substantial scatter in
the subhalo mass loss rates, which is well represented by a log-normal
with a standard deviation of $\sigma_{\log(\dot{m}/M)} \sim
0.25$. However, G08 measured the mass loss rates averaged over a time
step of $\sim 0.1$Gyr, which is much shorter than an orbital
period. Hence, their mass loss rates are not truly orbit-averaged, and
their scatter contains a contribution due to variations in $\dot{m}/M$
along individual orbits. In order to gauge the scatter in {\it
  orbit-averaged} subhalo mass loss rates due to the variance in
orbital properties and halo concentrations we consider a simple, but
insightful, toy model.

Consider a subhalo of mass $m$ on an orbit of energy $E$ and angular
momentum $L$ (both per unit mass) inside a host halo of mass $M$. In
what follows we use $r$ and $R$ to refer to halo-centric radii in the
subhalo and host halo, respectively.  As the subhalo orbits within the
potential of the host halo, it experiences dynamical friction, tidal
heating and tidal stripping, all of which contribute to the subhalo
loosing mass. Most of this mass loss occurs close to the orbit's
pericenter, $R_\rmp$, where the tidal field of the host is
strongest. Hence, we may approximate the orbit averaged mass loss rate
of a subhalo as
\begin{equation} \label{masslossmodel}
\dot{m} = {m - m(r_\rmt) \over T_\rmr}.
\end{equation}
where $r_\rmt$ is the subhalo's tidal radius at the orbit's
pericenter, $T_\rmr$ is the radial orbital period, and $m(r)$ is the
subhalo mass enclosed within radius $r$. Hence, we assume that all the
mass beyond the tidal radius of the subhalo at the orbit's pericenter
is stripped from the subhalo in one radial period. This is admittedly
a crude, and poorly justified, assumption, but as we will show below,
it yields results in close agreement with numerical simulations.
\begin{figure*}
\centerline{\psfig{figure=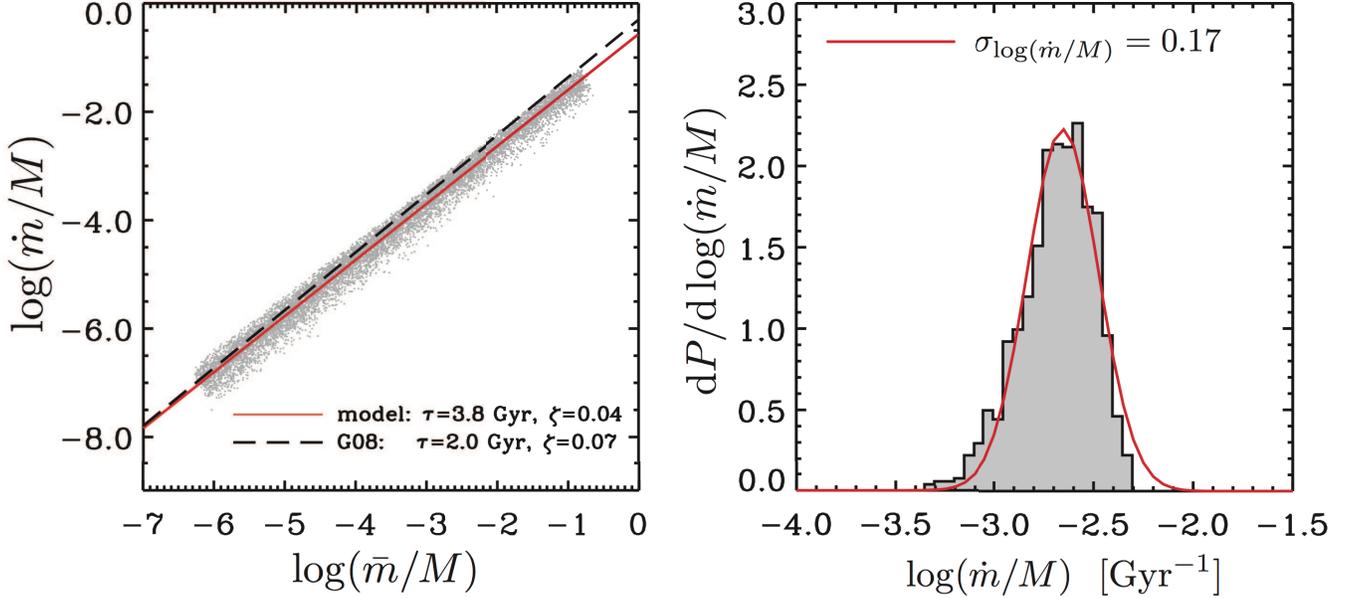,width=1.0\hdsize}}
\caption{Results obtained from the toy model.  {\it Left Panel}:
  orbit-averaged mass-loss rates, $\dot{m}$, as a function of the
  orbit-averaged subhalo mass, $\bar{m}$, for a host halo of mass $M =
  10^{13}\Msunh$ at $z=0$. Grey dots represent the 10,000 individual
  Monte-Carlo realizations. The solid red line is the best-fit to the
  median, which corresponds to Eq.(\ref{mydecay})-(\ref{mytau}) with
  $\calA = 0.81$ and $\zeta=0.04$.  The dashed black line represents
  the average subhalo mass-loss rates of G08, as measured from
  numerical simulations, which has $\calA = 1.54$ and $\zeta=0.07$.
  {\it Right Panel}: the distribution of orbit-averaged mass-loss
  rates at $\log(\bar{m}/M)=-2$.  The distribution can be approximated
  as a log-normal distribution with dispersion
  $\sigma_{\log(\dot{m}/M)} = 0.17$ (red curve).  }
\label{Fig:MassLossRate}
\end{figure*}

Throughout we assume that both host haloes and subhaloes are defined
as spheres with an average density, inside their virial radii, given
by $\bar{\rho}_\rmh = \Delta_{\rm crit}(z) \rho_{\rm crit}(z)$, and
with an NFW density profile (Navarro, Frenk \& White 1997). Hence,
their enclosed mass profile is
\begin{equation} \label{NFWmass}
M(R) = M \, {f(c \, R/R_{\rm vir}) \over f(c)}\,,
\end{equation}
with $c$ the halo concentration parameter, $R_{\rm vir}$ the halo
virial radius, and $f(x)=\ln(1+x)-x/(1+x)$. We assume that, at fixed
halo mass, the concentrations follow a log-normal distribution with
standard deviation $\sigma_{\log c} \simeq 0.12$ (e.g., Macci\`o \etal
2010) and with a {\it median} that depends on halo mass and redshift
according to
\begin{equation} \label{concmass}
c(M,z) = {4.67 \over 1+z} \, \left( {M(z) \over 10^{14}\Msunh}\right)^{-0.11}
\end{equation}
(Neto \etal 2007). 

With the mass profile of the host halo specified, we can determine the
apocenter, $R_\rma$ and pericenter, $R_\rmp$ of the subhalo's orbit,
by solving for the roots of
\begin{equation} \label{ApoPeri}
{1\over R^2} + {2[\Phi(R)-E] \over L^2}=0
\end{equation}
(Binney \& Tremaine 2008). Here
\begin{equation} \label{NFWpot}
\Phi(R) = -V^2_{\rm vir} \, {\ln(1+c \, R/R_{\rm vir}) \over f(c) \, R/R_{\rm vir}}
\end{equation}
is the gravitational potential of the NFW host halo, with $V_{\rm vir}
= \sqrt{G\,M/R_{\rm vir}}$ the host halo's virial velocity. The radial
orbital period is given by
\begin{equation} \label{Tr}
T_\rmr = 2\int_{R_\rmp}^{R_\rma} {\rmd R \over \sqrt{2[E-\Phi(R)]-L^2/R^2} }\,,
\end{equation}
(Binney \& Tremaine 2008), and the tidal radius of the subhalo at
$R=R_\rmp$ is obtained by solving
\begin{equation} \label{TidalRadius}
r_\rmt = R_\rmp \left[  { m(r_\rmt)/M(R_\rmp) \over 
2 + {\Omega_\rmp^2 \, R_\rmp^3 \over G \, M(R_\rmp)} -
{\rmd\ln M \over \rmd\ln R}\Huge\vert_{R_\rmp} }  \right]^{1/3}\,,
\end{equation}
(e.g., von Hoerner 1957; King 1962; Taylor \& Babul 2001), where
\begin{equation} \label{AngSpeed}
\Omega_\rmp = L/R_\rmp^2
\end{equation}
is the instantaneous angular speed at pericenter. 

The final ingredient for our toy model is the probability distribution,
$\calP(E,L)$, for the orbital energies and angular momenta of dark
matter subhaloes. For convenience, we characterize $E$ and $L$ via the
radius of a circular orbit, $R_\rmc$, and the orbital circularity, $\eta$.
The relations between $(E,L)$ and $(R_\rmc,\eta)$ are given by
\begin{equation} \label{OrbitalEnergy}
E = {1 \over 2}V_\rmc^2 + \Phi(R_\rmc)\,,
\end{equation}
with $V_\rmc$ the circular speed at $R=R_\rmc$, and
\begin{equation} \label{DefineCircularity}
L = \eta \, L_\rmc(E)
\end{equation}
with $L_\rmc(E) = R_\rmc\,V_\rmc$ the maximum angular momentum for an
orbit of energy $E$.  Using high-resolution numerical simulations,
Zentner \etal (2005) has shown that the circularity distribution for
the orbits of subhaloes at infall is well fit by
\begin{equation} \label{Pcirc}
P(\eta) \propto \eta^{1.22} \, (1-\eta)^{1.22}\,.
\end{equation}
Zentner \etal (2005) also showed that the distribution of $R_\rmc$
for the infalling subhaloes is well approximated by a uniform
distribution covering the range $[0.6\,R_{\rm vir},R_{\rm vir}]$,
i.e.,
\begin{equation} \label{PRc}
\calP(R_\rmc) = \left\{ 
\begin{array}{ll}
  5/3 & \mbox{if $0.6 \leq R_\rmc/R_{\rm vir} \leq 1.0$} \\
   0  & \mbox{otherwise}
\end{array}\right.
\end{equation}
For our toy model we follow Gan \etal (2010) and assume that the
probability distribution for $R_\rmc$ and $\eta$ is separable, i.e.,
$\calP(R_\rmc,\eta) = \calP(R_\rmc) \, \calP(\eta)$, and draw the
values for $R_\rmc$ and $\eta$ from Eqs.~(\ref{PRc})
and~(\ref{Pcirc}), respectively\footnote{Actually, we sample the
  circularities from the modified distribution, $\calP(\eta)\,
  \rmd\eta = {\pi\over 2}\, \sin(\pi\eta) \, \rmd\eta$, which
  accurately fits Eq.~(\ref{Pcirc}) and has the advantage that it
  allows values of $\eta$ to be drawn by direct inversion.}.

Using this toy model, we compute orbit-averaged mass loss rates for
large ensembles of subhaloes as follows. For a given host halo mass,
we first draw a subhalo mass, $m$, from a uniform distribution on the
interval $\log(m/M) \in [-6.0,-0.5]$. \footnote{We have verified that
  drawing subhalo mass from the unevolved SHMF instead does not alter
  any of the results.}  Next we draw concentrations for both the host
halo and subhalo, using the log-normal distribution described above,
as well as values for the orbital energy and angular momentum of the
subhalo. These are used to compute the radial orbital period,
$T_\rmr$, and the subhalo's tidal radius, $r_{\rm t}$, from which we
ultimately compute the mass loss rate using
Eq.~(\ref{masslossmodel}). The left-hand panel of
Fig.~\ref{Fig:MassLossRate} shows the resulting orbit-averaged mass
loss rates, $\dot{m}/M$, as function of the orbit averaged subhalo
mass, $\bar{m}/M$, where $\bar{m} = m - (\dot{m}\,T_\rmr)/2$. Both
have been normalized by the host halo mass, $M$, for convenience. The
dots are the results obtained for our toy model, using 10,000
subhaloes for a host halo of mass $M = 10^{13} \Msunh$. We find that
our model results are well fit by Eqs.~(\ref{mydecay})-(\ref{mytau})
with $\calA = 0.81$ and $\zeta = 0.04$, which is shown as the solid
line. Hence, our toy model lends further support to the functional
form of the average subhalo mass loss rate introduced by B05. For
comparison, the dashed line is the best-fit relation obtained by G08
from their numerical simulations. As is apparent, the latter is
somewhat steeper ($\zeta = 0.07$), and implies somewhat larger mass
loss rates ($\calA = 1.54$). We believe that this discrepancy between
our toy model and the G08 simulation results is most likely due to the
fact that our toy model ignores dynamical friction, which reduces the
orbit's pericentric distance as well as the radial orbital period,
both of which will boost the mass loss rate.  Since dynamical friction
is more important for more massive subhaloes, taking dynamical
friction into account is likely to increase both $\zeta$ and $\calA$,
bringing the toy model in better agreement with the simulation
results.

As mentioned above, the main goal of the toy model is to gain some
insight regarding the scatter in $\dot{m}/M$ that arises due to
variance in the orbital properties and halo concentrations. The right
panel of Fig.~\ref{Fig:MassLossRate} plots the distribution of
normalized subhalo mass loss rates, $\rmd\calP/\rmd(\dot{m}/M)$, at
fixed $\bar{m}/M = 0.01$, which is well fit by a log-normal with a
standard deviation $\sigma_{\log(\dot{m}/M)} = 0.17$ (red curve). Note
that this scatter is substantially smaller than the
$\sigma_{\log(\dot{m}/M)} \simeq 0.25$ measured by G08 from their
simulations, as expected. In fact, our results imply that the
contribution to the variance measure by G08 due to variations in
$\dot{m}/M$ {\it along} an orbit are of the order of $\sqrt{0.25^2 -
  0.17^2} \simeq 0.18$, comparable to the scatter that arises from
variations in orbital properties, which is roughly what one
expects. Since our model requires orbit-averaged mass loss rates, we
will model the scatter in $\dot{m}/M$ using a log-normal distribution
with $\sigma_{\log(\dot{m}/M)} = 0.17$ (see \S\ref{Sec:compute}
below).

\subsection{Converting Mass to Maximum Circular Velocity}
\label{Sec:Vmax}

In addition to subhalo mass, we also want the semi-analytical model to be
able to yield the maximum circular velocity $V_{\rm max}$ for each of
its subhaloes.  The maximum circular velocity of a halo of mass $M$
depends on its density distribution. In the case of NFW haloes
\begin{equation} \label{VmaxHost}
V_{\rm max} = 0.465 \, V_{\rm vir} \, \sqrt{c \over \ln(1+c) - c/(1+c)}\,,
\end{equation}
where
\begin{eqnarray}\label{Vvir}
\lefteqn{V_{\rm vir} = 159.43 \kms \, \left({M \over 10^{12}\msunh}\right)^{1/3}
\, \left[{H(z) \over H_0}\right]^{1/3}} \nonumber \\
& & \, \left[{\Delta_{\rm vir}(z) \over 178}\right]^{1/6}\,,
\end{eqnarray}
is the virial velocity of a dark matter halo of virial mass $M$ at
redshift $z$. It is well known that the concentration of a dark matter
halo is strongly correlated with its MAH, in the sense that haloes
that assemble earlier are more concentrated (e.g., Navarro \etal 1997;
Wechsler \etal 2002; Ludlow \etal 2013). We use the model of Zhao
\etal (2009), according to which
\begin{equation}
\label{cvir}
c(M,t) = 4.0\, \left[1+\left( {t \over 3.75\,t_{0.04}} \right)^{8.4} \right]^{1/8}.
\end{equation}
Here $t_{0.04}$ is the proper time at which the host halo's main
progenitor gained 4 percent of its mass at proper time $t$, which we
extract from the halo's merger tree, as described below.
\begin{figure}
\centerline{\psfig{figure=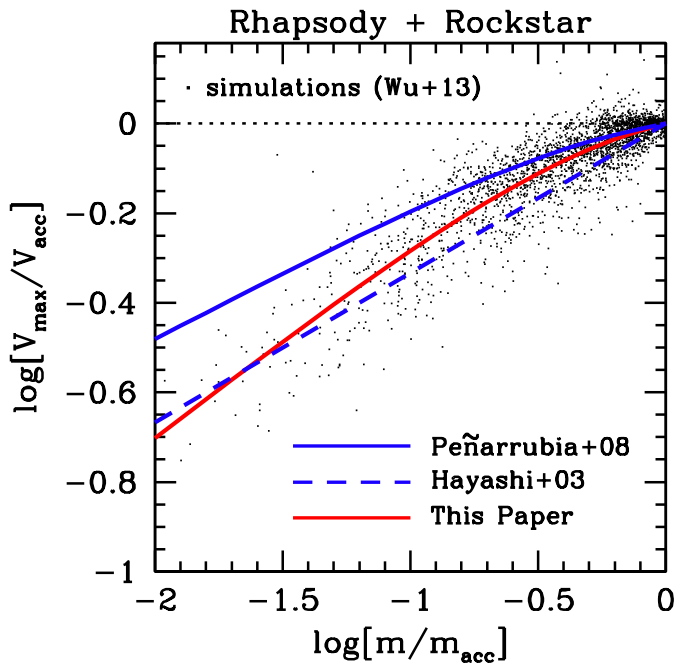,width=0.5\hdsize}}
\caption{The ratio $V_{\rm max}/V_{\rm acc}$ as function of $m/m_{\rm
    acc}$ for 2735 subhaloes in the Rhapsody simulations (black
  dots). The solid, red line is the best-fit equation of the
  form~(\ref{Vmaxfit}) with best-fit parameter $(\eta,\mu) =
  (0.44,0.60)$. This is the relation we use throughout to compute the
  maximum circular velocities of dark matter subhaloes in our
  semi-analytical model. The solid and dashed blue curves are the best-fit
  results of Hayashi \etal (2003) and Pe\~narrubia \etal (2010), and
  are shown for comparison.}
\label{Fig:VmaxMsubRelation}
\end{figure}

For subhaloes, we use the results of Hayashi \etal (2003) and
Pe\~{n}arrubia \etal (2008, 2010), who, using idealized numerical
simulations, have shown that the evolution of the maximum circular
velocity of a dark matter subhalo depends solely on the total amount
of mass stripped, and not on the details of how or when that mass is
stripped. Following Pe\~{n}arrubia \etal (2008) we therefore write
\begin{equation}\label{Vmaxfit}
V_{\rm max} = 2^{\mu} \, V_{\rm acc} \, {(m/m_{\rm acc})^{\nu} \over
 (1 + m/m_{\rm acc})^{\mu}}\,,
\end{equation}
where $m_{\rm acc}$ and $V_{\rm acc}$ are the subhalo mass and maximum
circular velocity at the time of accretion. In order to constrain the
two free parameter $\nu$ and $\mu$, we use results from the Rhapsody
simulation project (Wu \etal 2013a,b), a large suite of 96
high-resolution, simulations of cluster-sized dark matter haloes with
a present-day mass of $M = 10^{14.8\pm 0.05}\Msunh$.  These have been
re-simulated with high resolution from a cosmological volume of $1
h^{-3} {\rm Gpc}^3$ in a $\Lambda$CDM cosmology with $\Omega_\rmm =
0.25$, $\Omega_{\Lambda} = 0.75$, $\Omega_\rmb = 0.04$, $h=0.7$,
$\sigma_8 = 0.8$ and spectral index $n_\rms = 1.0$ (hereafter
`Rhapsody cosmology').  As in our model, the host haloes are defined
as spherical, overdense regions with an average density equal to
$\Delta_{\rm vir}\rho_{\rm crit}$. Using the 6D phase-space halo
finder {\tt ROCKSTAR} (Behroozi \etal 2013a,b), Wu \etal (2013b)
measured subhalo mass and velocity functions covering more than four
orders of magnitude in subhalo mass. Fig.~\ref{Fig:VmaxMsubRelation}
plots the ratio $V_{\rm max}/V_{\rm acc}$ as function of $m/m_{\rm
  acc}$ for a random subset of 2735 subhaloes from the Rhapsody
simulations, kindly provided to us in electronic format by
H. Wu. Fitting Eq.~(\ref{Vmaxfit}) to these data we obtain the
best-fit relation indicated by the red curve, which has
$(\nu,\mu)=(0.44,0.60)$. As we show in Paper~II, the same relation
also accurately fits results from the Bolshoi (Klypin \etal 2011) and
MultiDark (Prada \etal 2012; Riebe \etal 2013) simulations, and is
independent of host halo mass.  For comparison, the solid and dashed
blue curves are the $V_{\rm max}/V_{\rm acc}$ - $m/m_{\rm acc}$
relations obtained by Pe\~{n}arrubia \etal (2010) and Hayashi \etal
(2003), respectively, using high-resolution, idealized $N$-body
simulations of individual subhaloes orbiting in a static, spherical
NFW host halo.  These roughly bracket the results from the
cosmological Rhapsody simulations.

\subsection{Computing Subhalo Mass and Velocity Functions}
\label{Sec:compute}

Having described all the ingredients, we now outline how these are
combined to compute the subhalo mass and velocity functions of a host
halo of mass $M_0$ at redshift $z_0$. We first construct a merger
tree, with a mass resolution of $\psi_{\rm res} = 10^{-5}$, as
described in \S\ref{Sec:Trees}. Next, we follow each trajectory
forward in time, starting from the redshift $z_{\rm acc}$ at which the
trajectory's halo first becomes a subhalo, evolving the subhalo mass
in between merger-tree time steps all the way to $z=z_0$. The subhalo
mass loss rate is given by Eqs.~(\ref{mydecay}) - (\ref{mytau}) with
$\zeta = 0.07$, and a normalization $\calA$ that is a
trajectory-specific random variable drawn from the following
log-normal distribution
\begin{equation}\label{ProbA}
\calP(\calA) \, \rmd\calA = {\log e \over \sqrt{2\pi} \scatter} \,
\exp\left[-{\log^2(\calA/\bar{\calA}) \over 2 \scatter^2}\right] \, 
{\rmd\calA \over \calA}\,.
\end{equation}
Here $\scatter = \sigma_{\log(\dot{m}/M)} = 0.17$ and the median
$\bar{\calA}$ is our model's single free parameter, which we tune to
reproduce the $z=0$ subhalo mass function from simulations as
described in \S\ref{Sec:Test1st} below. Note that this scatter in the
normalization $\calA$ takes account of the scatter in subhalo mass
loss rates due to the variance in orbital properties and halo
concentrations.

Note that we loop over trajectories sorted by increasing order, which
assures that each subhalo is evolved in its properly evolved parent
halo. Throughout we also assume that subhaloes always continue to
orbit within the parent halo that directly hosts it: in particular, if
a first-order sub-halo of mass $m_1$ is orbiting within a zeroth-order
parent halo of mass $m_0$, and hosts a second-order sub-halo (i.e., a
sub-subhalo) of mass $m_2$, the latter is evolved using
Eq~(\ref{mydecay}) with $M=m_1$ and $m=m_2$, while $m_1$ is
evolved using the same equation but with $M=m_0$ and $m=m_1$. In other
words, we ignore the possibility that a higher-order subhalo is
stripped from its direct parent, which would cause its order to
decrease by one. We also ignore the possibility of subhalo-subhalo
mergers inside a parent halo, which would cause the order of the
less-massive subhalo to increase by one. As discussed in Paper~II,
these oversimplified assumptions do not seem to have a significant
impact on the accuracy of our model.

In order to compute the subhalo velocity function, $\rmd
N/\rmd\log(V_{\rm max}/V_{\rm vir})$, with $V_{\rm vir}$ the virial
velocity of the host halo, we compute $V_{\rm max}$ at each time step
along each trajectory of the host halo's merger tree using the
following approach. Starting from the trajectory's leave point, we
first use the P08 algorithm to construct the mass accretion history
(MAH) back in time until the leaf's most massive progenitor reaches a
mass $M < 0.04 M_{\rm leaf}$. This `extension' of the merger tree is
used to compute $t_{0.04}$ (using simple linear interpolation in
between time steps), which is used in turn to compute $V_{\rm max}$
for the leaf halo using Eqs.~(\ref{VmaxHost}) - (\ref{cvir}). Tracing
the trajectory forward in time, each time step we use the past
trajectory (plus its extension) to compute $V_{\rm max}$ using the
same approach. Once the halo associated with the trajectory becomes a
subhalo, it starts to experience mass loss, as described by
Eq.~(\ref{mydecay}), and we use Eq.~(\ref{Vmaxfit}) with $(\nu,\mu) =
(0.44,0.60)$ to compute the evolution of its maximum circular
velocity. As we demonstrate in \S\ref{Sec:Test} below, this method
yields SHVFs in excellent agreement with simulation results (see also
Paper~II).


\section{Results}
\label{Sec:Test}

Our model as described above yields, at each redshift, the evolved
SHMF and SHVF for all orders of subhaloes.  In this section, we
tune the free parameter $\bar{\calA}$, compare the resulting model
predictions with a few simulation results, and discuss how the
subhalo mass and velocity functions scale with halo mass and redshift.

\subsection{Testing and Calibrating the Model}
\label{Sec:Test1st}

\begin{figure}
\centerline{\psfig{figure=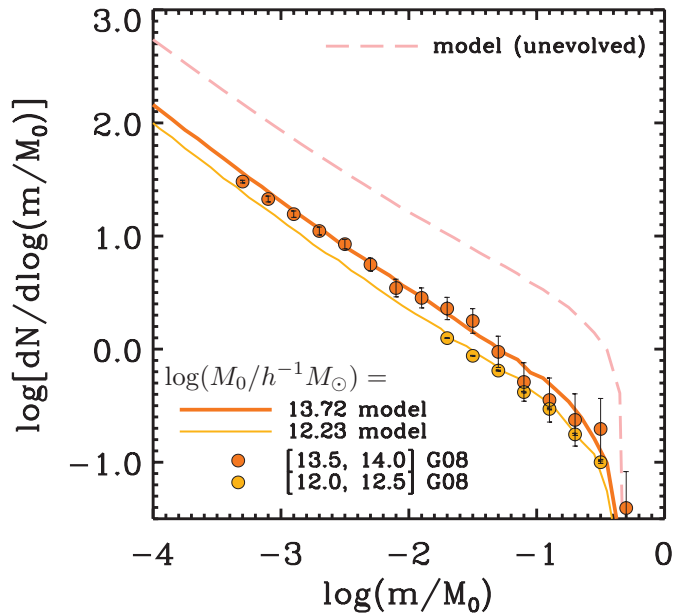,width=0.5\hdsize}}
\caption{Comparison of average, evolved SHMFs at $z=0$ obtained from
  $N$-body simulations (symbols with errorbars) with predictions from
  our semi-analytical model (solid lines). The simulation data is taken
  from the study by G08 and corresponds to host halo mass bins of
  $\log[M_0/(\Msunh)] \in [12.0,12.5]$ and $[13.5,14.0]$, as
  indicated. The model results have been obtained averaging over
  10,000 Monte Carlo realizations for host haloes of $M_0 =
  10^{12.23}h^{-1}\Msun$ and $10^{13.72}h^{-1}\Msun$, which are the
  average halo masses in the corresponding mass bins from the
  simulation.  Finally, the dashed curve is the unevolved subhalo mass
  function, obtained from our merger trees, and is shown for
  comparison.}
\label{Fig:ESMF_CompareWithG08}
\end{figure}

The symbols with errorbars in Fig.~\ref{Fig:ESMF_CompareWithG08} are
the evolved SHMFs of first-order subhaloes at $z_0 = 0$, obtained by
G08 using a set of simulations for a flat
$\Lambda$CDM cosmology with $\Omega_\rmm = 0.3$, $\Omega_{\Lambda} =
0.7$, $\Omega_{\rmb} = 0.04$, $h = H_0/(100 \kmsmpc) = 0.7$ and with
initial density fluctuations described by a scale-invariant power
spectrum with normalization $\sigma_8=0.9$. Host haloes are defined as
spheres with an average density equal to $\Delta_{\rm vir} \,
\rho_{\rm crit}$, and their subhaloes are identified using the subhalo
finder {\tt SURV}\footnote{{\tt SURV} differs from most other subhalo
  finders in that it uses prior information based on the host halo's
  merger tree to identify its subhaloes.} that was developed by Tormen
\etal (2004) and G08. Results are shown for two mass
bins: $\log[M_0/(h^{-1}\Msun)] \in [12.0,12.5]$, for which the average
is $12.23$, and $\log[M_0/(h^{-1}\Msun)] \in [13.5,14.0]$, for which
the average is $13.72$. These bins contain a total of 3349 and 127
host haloes, respectively. The error bars indicate the rms values of 
$\rmd N/\rmd\log(m/M_0)$ in the logarithmic $m/M_0$-bins with the 
width of 0.2 dex and bin-centers corresponding to the positions of the symbols.

The solid curves in Fig.~\ref{Fig:ESMF_CompareWithG08} are the
corresponding model predictions for host haloes of mass $M_0 =
10^{12.23}\Msunh$ and $M_0 = 10^{13.72}\Msunh$, respectively, obtained
by averaging over 10,000 Monte Carlo realizations (i.e., merger
trees). We have tuned $\bar{\calA}$, which basically controls the
overall normalization, such that the model curves match the G08
simulation results. The resulting best-fit value is $\bar{\calA} =
1.34$ (corresponding to a present-day mass-loss time scale of $\tau_0
= 2.3$Gyr). Interestingly, with $\bar{\calA}$ close to unity, we have
that the average time scale for subhalo mass loss is basically just
the dynamical time of the halo. We return to this somewhat intriguing
result in \S\ref{Sec:Discussion}. From here on we keep $\bar{\calA}$
fixed, so that the model is fully determined.

Note that our best-fit value for $\bar{\calA}$ is in good agreement
with the $\bar{\calA} = 1.54^{+0.52}_{-0.31}$ obtained by G08 from the
median mass loss rates inferred directly from their simulations. This
is a dramatic improvement with respect to B05, who inferred
$\bar{\calA} = 23.7$, and owes entirely to the use of more
accurate merger trees. Note that our model accurately reproduces the
slope and host mass dependence of SHMF (see also Paper~II and
\S\ref{Sec:MassDep} below), which is a feature that cannot be
controlled by the freedom in $\bar{\calA}$.  It is worth mentioning
that the stochasticity in the mass-loss model ($\scatter = 0.17$)
actually causes a boost of the normalization of the SHMF; if we ignore
this stochasticity, the best-fit value for the median mass-loss
normalization is $\bar{\calA} \sim 1.2$. The fact that the best-fit
value for $\bar{\calA}$ depends on the (amount of) stochasticity used
is a manifestation of the log-normal nature of the distribution of
orbit-averaged mass loss rates (see right-hand panel of
Fig.~\ref{Fig:MassLossRate}), which causes the mean to be larger than
the median.

Finally, the dashed curve in Fig.~\ref{Fig:ESMF_CompareWithG08}
represents the {\it unevolved} SHMFs obtained from the same Monte
Carlo merger trees. Note that the unevolved SHMF is independent of
host halo mass (see e.g., B05; Li \& Mo 2009; Yang \etal 2011), and is
therefore identical for the two mass bins shown. The differences
between this (universal) dashed curve and the solid curves reflects
the global impact of subhalo mass stripping integrated over the
assembly history of the host halo.

\begin{figure*}
\centerline{\psfig{figure=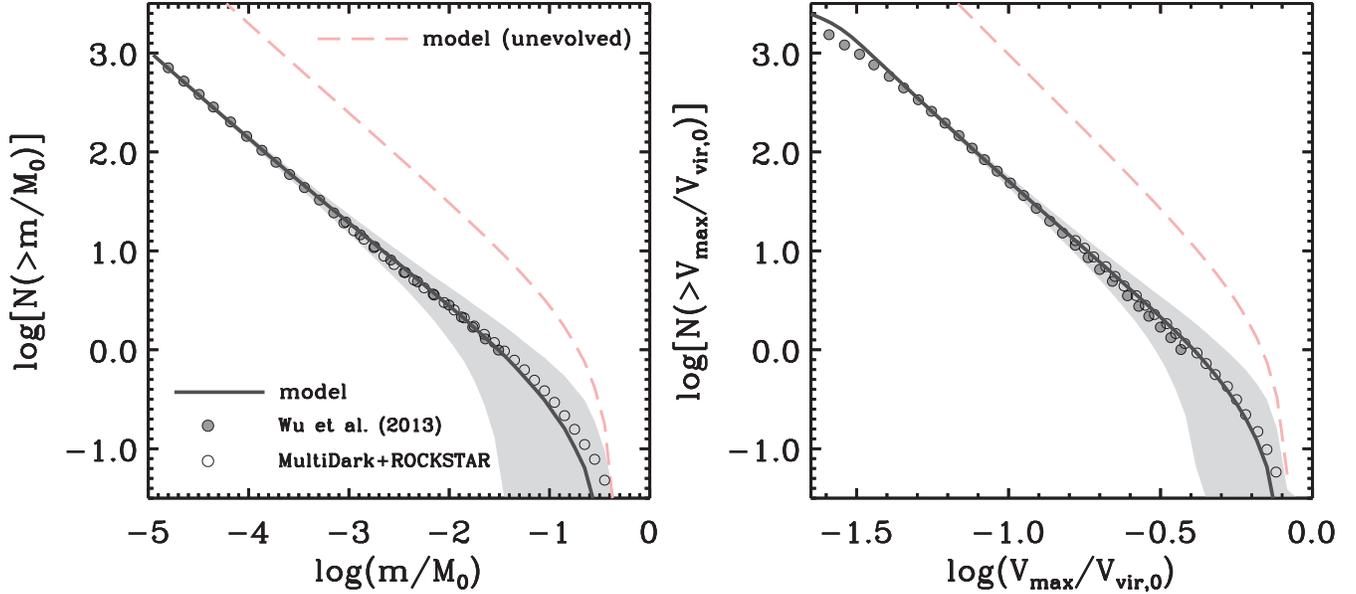,width=1.0\hdsize}}
\caption{The average, cumulative mass function (left-hand panel) and
  velocity function (right-hand panel) at $z=0$ for subhaloes of all
  orders in host haloes of mass $M_0 \simeq 10^{14.8}\Msunh$. The
  filled and open circles are results from the high-resolution
  Rhapsody simulations (Wu \etal 2013b; $M_0 = 10^{14.8\pm
    0.05}\Msunh$) and the publicly available MultiDark simulation
  (Prada \etal 2012; $M_0\in[10^{14.5},10^{15}]\Msunh$), respectively,
  and have both been obtained using the (sub)halo finder {\tt
    ROCKSTAR}. The solid curves are our model predictions for $M_0 =
  10^{14.8}\Msunh$, averaged over 10,000 Monte Carlo realizations
  based on the Rhapsody cosmology. The gray bands indicate the
  corresponding halo-to-halo variance, while the dashed curves
  indicate the corresponding unevolved mass and velocity functions.
  Note the exquisite agreement between model and simulation results.}
\label{Fig:ESMF_CompareWithWu13}
\end{figure*}

To further test our model, we once more use the Rhapsody simulations
of Wu \etal (2013a,b). The filled circles in
Fig.~\ref{Fig:ESMF_CompareWithWu13} are the average, cumulative mass
and velocity functions for subhaloes of all orders  obtained by
Wu et al., using the subhalo finder {\tt ROCKSTAR}. Note that these
are normalized with respect to the present day mass, $M_0$, and virial
velocity, $V_{{\rm vir},0}$, of their host halo.  Because of the high
resolution of the Rhapsody simulations, the Wu \etal data probes
subhaloes all the way down to $m = 10^{-5} M_0$ ($V_{\rm max} \sim
0.03 V_{{\rm vir},0}$), an improvement of $\sim 1.5$ orders of
magnitude in normalized subhalo mass with respect to the G08 results
shown in Fig.~\ref{Fig:ESMF_CompareWithG08}. However, at the massive
end the results published by Wu \etal (2013b) only extent to $m \sim
0.03 M_0$ ($V_{\rm max} \sim 0.3 V_{{\rm vir},0}$), which is roughly
where the cumulative abundances drop below unity.  We therefore
complement the data of Wu \etal (2013b) with simulation data from the
MultiDark simulation, which covers a cosmological volume of $1 h^{-3}
{\rm Gpc}^3$, but at significantly lower resolution.  Using the
publicly available {\tt ROCKSTAR}
catalog\footnote{http://hipacc.ucsc.edu/Bolshoi/MergerTrees.html} of
haloes and subhaloes at $z=0$, we compute the cumulative SHMF and
SHVF, averaged over 2393 host haloes with masses in the range
$10^{14.5} \Msunh \leq M_0 \leq 10^{15} \Msunh$.  The results are
shown as open circles in Fig.~\ref{Fig:ESMF_CompareWithWu13}, and
cover the ranges $m/M_0 \gta 10^{-3}$ and $V_{\rm max}/V_{{\rm vir},0}
\gta 0.15$. Note that the MultiDark and Rhapsody results are in
excellent agreement in the range where they overlap.  This is true
despite the fact that the MultiDark simulation corresponds to a
slightly different cosmology ($[\Omega_\rmm,
  \Omega_{\Lambda},\Omega_\rmb,h,\sigma_8,n_\rms] = [0.27, 0.73,
  0.047, 0.7, 0.82, 0.95]$) than the Rhapsody simulations. Using our
semi-analytical model, we have verified though that this slight difference
in cosmological parameters has a negligible impact on the SHMF (see
Paper~II).

The solid curves in Fig.~\ref{Fig:ESMF_CompareWithWu13} correspond to
our model predictions, averaged over 10,000 Monte Carlo realizations
of host haloes with a present-day mass of $M_0=10^{14.8}\Msunh$ in the
Rhapsody cosmology. In order to estimate the SHMF and SHVF for
subhaloes of all orders we have summed the contributions of orders one
to four; as shown below, the contribution of subhaloes of even higher
order is negligible (see also JB14). The grey bands indicate the
standard deviation due to halo-to-halo variance, while the dashed
curves reflect the {\it unevolved}, cumulative SHMF and SHVF.  Note
that our semi-analytical model accurately reproduces the simulation results
over the entire ranges in $m/M_0$ and $V_{\rm max}/V_{{\rm vir},0}$
shown, without tuning any of the parameters in our mass loss model or
in the P08 merger tree algorithm.  Hence, we conclude that our model
can accurate reproduce both the SHMFs and SHVFs for subhaloes of first
order and all orders, and for different host halo masses and
cosmologies. This conclusion is further strengthened in Paper~II,
where we compare our model predictions with an even larger set of
simulation results.

\subsection{Mass and Redshift Dependence}
\label{Sec:MassDep}

Having tested and calibrated our mass-loss model, we now explore how
the subhalo mass and velocity functions scale with host halo mass,
redshift and subhalo order. To that extent we compute the average
SHMFs, $\rmd N/\rmd\log(m/M_0)$, and SHVFs, $\rmd N/\rmd\log(V_{\rm
  max}/V_{{\rm vir},0})$, averaged over 10,000 Monte Carlo
realizations, for host haloes of masses $\log[M_0/(h^{-1}\Msun)] = 11,
12,..,15$, and redshifts $z_0 = 0, 1, 3$ and $5$ in the Rhapsody
cosmology.

\begin{figure*}
\centerline{\psfig{figure=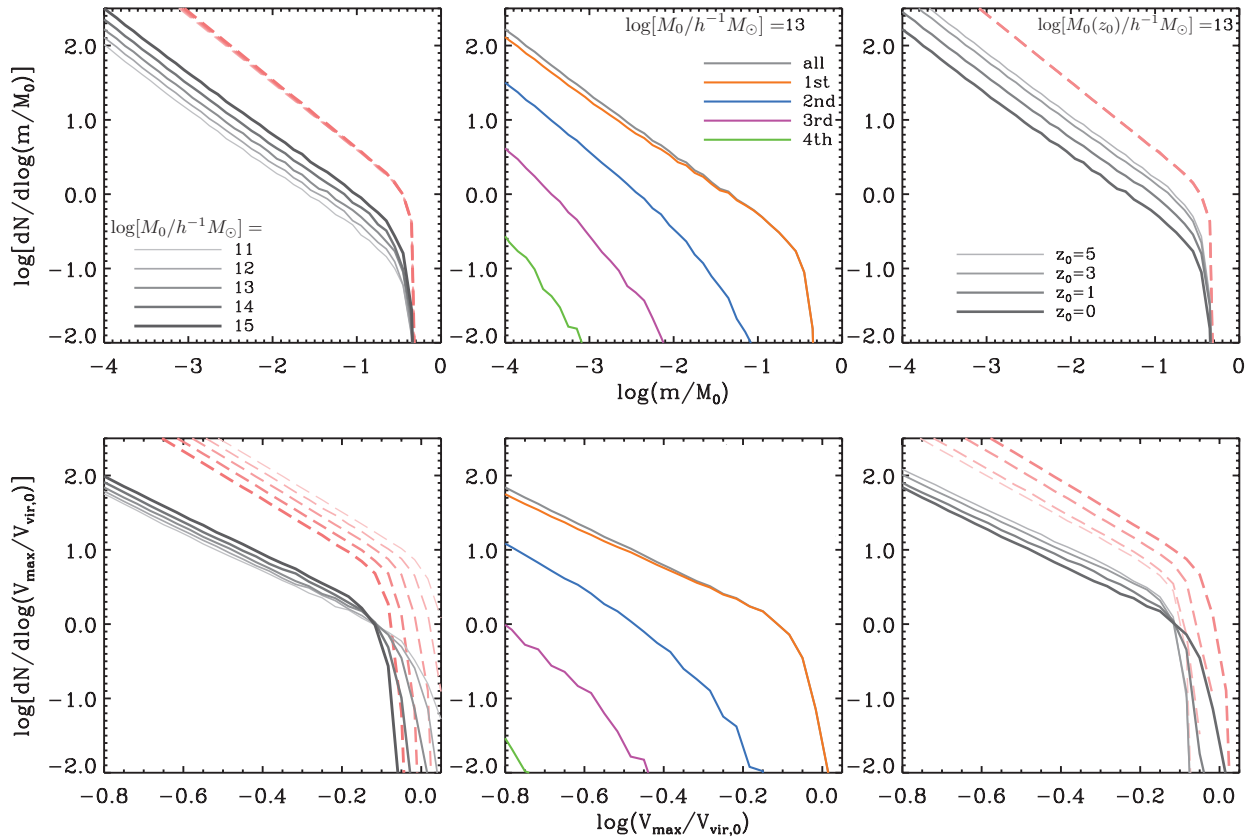,width=0.93\hdsize}}
\caption{Average SHMFs (upper panels) and SHVFs (lower panels)
  obtained by averaging over 10,000 merger trees each. {\it Left-hand
    panels:} Results for different host halo masses, as indicated,
  with solid and dashed curves corresponding to the evolved and
  unevolved mass/velocity functions, respectively. {\it Middle
    panels:} Evolved SHMFs and SHVFs of different orders (different
  colors, as indicated), for a host halo of mass $M_0 = 10^{13}
  \Msunh$ at $z=0$. {\it Right-hand panels:} Results for a host halo
  of mass $M_0 = 10^{13} \Msunh$ at different redshifts, as
  indicated. As in the left-hand panels, solid and dashed curves
  correspond to evolved and unevolved mass/velocity functions,
  respectively. Note how the unevolved SHMF is independent of mass and
  redshift, which is not the case for the unevolved SHVF.}
\label{Fig:MassDep}
\end{figure*}

The upper, left-hand panel of Fig.~\ref{Fig:MassDep} plots the $z=0$
SHMFs for the five different host halo masses, as indicated. Solid and
dashed curves correspond to evolved and unevolved SHMFs. As already
mentioned several times, the latter is universal, and therefore
displays no mass dependence. The evolved SHMFs on the other hand, are
clearly mass-dependent with a normalization that increases
systematically with increasing host halo mass. This mass dependence,
first noticed in numerical simulations by Gao \etal (2004) and in
semi-analytical models by B05, simply reflects that more massive
haloes form later: the universality of the unevolved SHMF shows that,
on average, all haloes accrete (sub)haloes of the same, normalized
mass. Since subhaloes that are accreted earlier will be more depleted,
host haloes that assemble earlier, accrete their subhaloes earlier,
which thus will be more depleted by the present day.

The upper, middle panel of Fig.~\ref{Fig:MassDep} plots the $z=0$
SHMFs of different orders for a host halo of mass $M_0 = 10^{13}
\Msunh$, as indicated. Note how with each increasing order the
normalization reduces by roughly an order of magnitude, while the
slope steepens.  Hence, when computing the SHMF for subhaloes of all
orders one only has to sum the contributions of subhaloes of orders
one and two, unless one aims to reach exquisite precision. Throughout
this paper we always sum subhaloes up to fourth order, but none of our
results would change noticeably if we were to ignore the third and
fourth order subhaloes.  The gray line shows the SHMF of all orders,
which is clearly completely dominated by first-order subhaloes. Note,
though, that the slope of the SHMF of all orders is slightly, but
significantly, steeper than that of first-order subhaloes (see
\S\ref{Sec:Universal} below).

The upper, right-hand panel Fig.~\ref{Fig:MassDep} plots the SHMFs of
all orders for host haloes of mass $M_0 = M_0(z_0) = 10^{13} \Msunh$
at different redshifts $z_0$ as indicated.  At higher redshifts host
haloes of the same mass have a larger abundance of subhaloes than
their counterparts at lower redshifts. The unevolved SHMF, on the
other hand, is independent of redshift. As discussed in B05, the
subhalo mass fraction of a given halo at redshift $z$ is a trade-off
between the time scale, $\tau_{\rm acc}$, on which new subhaloes are
being `accreted' by the host halo, and the time scale, $\tau =
\tau_{\rm dyn}/\calA \simeq \tau_{\rm dyn}$, of subhalo mass loss.
The latter evolves with redshift as described by Eq.~(\ref{mytau}),
and therefore was shorter in the past.  The former depends on the
detailed mass assembly history of the host halo, and is thus a
function of both redshift and halo mass.  In the limit where
$\tau_{\rm acc} \ll \tau_{\rm dyn}$, subhalo mass loss is negligible
and the normalization of the SHMF will increase with time.  The
opposite limit, in which $\tau_{\rm acc} \gg \tau_{\rm dyn}$, is
equivalent to that of subhalo mass loss in a static parent halo.  In
this case, the normalization of the SHMF will decrease with
time. Since $\tau_{\rm acc}$ is of order the Hubble time, which is
always larger than the dynamical time of a halo, we are in the
latter regime, which explains why the normalization of the SHMF
decreases with decreasing redshift.

The lower panels of Fig.~\ref{Fig:MassDep} plot the same results as
the upper panels, but for the SHVFs. Overall the trends are very
similar except for one important difference: the unevolved SHVFs are
{\it not} universal. As is evident from the lower left- and lower
right-hand panels, the unevolved SHVF increases with decreasing host
halo mass and decreasing redshift. As discussed in
\S\ref{Sec:Universal} below, this is a consequence of the
concentration-mass-redshift relation of dark matter haloes, and causes
a cross-over in the mass- and redshift- dependence of the evolved
SHVFs at the massive end.


\section{Universal Models for the Subhalo Mass and Velocity Functions} 
\label{Sec:Universal}

Fig.~\ref{Fig:MassDep} suggests that the SHMF has a universal shape,
with a normalization that depends on halo formation time.  This
universality has its origin in the universal shape of the {\it
  unevolved} SHMF, and suggest that it should be straightforward to
obtain a fitting function for the {\it evolved} SHMF, $\rmd
N/\rmd\ln(m/M_0)$, that is valid for any $M_0$, any redshift, $z_0$,
and any cosmology. In fact, the results of Fig.~\ref{Fig:MassDep}
suggest that a similar universal fitting function may be found for
both the evolved and unevolved SHVFs.  In this section we present such
universal fitting functions, and demonstrate how their normalizations
can be computed from very simple considerations. The results of this
section are summarized in Appendix~A, where we provide a simple
step-by-step description of how to compute these universal fitting
functions.

Before we proceed, though, we caution that what follows should only be
used to describe subhalo mass and velocity functions in $\Lambda$CDM
cosmologies with parameters that are roughly consistent (within a
factor $\sim 2$) with current constraints. The reason is that the
alleged universality of the {\it unevolved} SHMF, first eluded to by
B05 and then confirmed in numerical simulations by G08 and Li \& Mo
(2009), only holds approximately. This was demonstrated in Yang \etal
(2011), who showed that the slope of the unevolved SHMF depends
weakly, but significantly, on the effective slope of the matter power
spectrum.

In order to have a sufficient dynamic range to probe how the subhalo
mass and velocity functions scale with mass and cosmology, we
construct average SHMFs and SHVFs for host haloes spanning the mass
range $10^{11} \Msunh \leq M_0 \leq 10^{15} \Msunh$ in cosmologies
that span the range $0.1 \leq \Omega_{\rmm} \leq 0.5$, $0.5 \leq
\sigma_8 \leq 1.0$ and $0.5 \leq h \leq 1.0$. Our `baseline' is a host
halo of mass $M_0 = 10^{13} \Msunh$ in the Rhapsody cosmology
[$(\Omega_{\rmm}, h, \sigma_8)=(0.25, 0.73, 0.8)$], which falls
roughly midway of the ranges considered. When varying cosmology, we
only vary one of these three cosmological parameters per time with
respect to this baseline cosmology and compute the average SHMF and
SHVF for host haloes with $M_0 = 10^{11} \Msunh$ and $10^{15}\Msunh$,
in each case averaging over 20,000 Monte Carlo realizations at
redshift $z_0 = 0$.

\subsection{Evolved Subhalo Mass Function} 
\label{Sec:EvolvedSHMF}

\begin{figure*}
\centerline{\psfig{figure=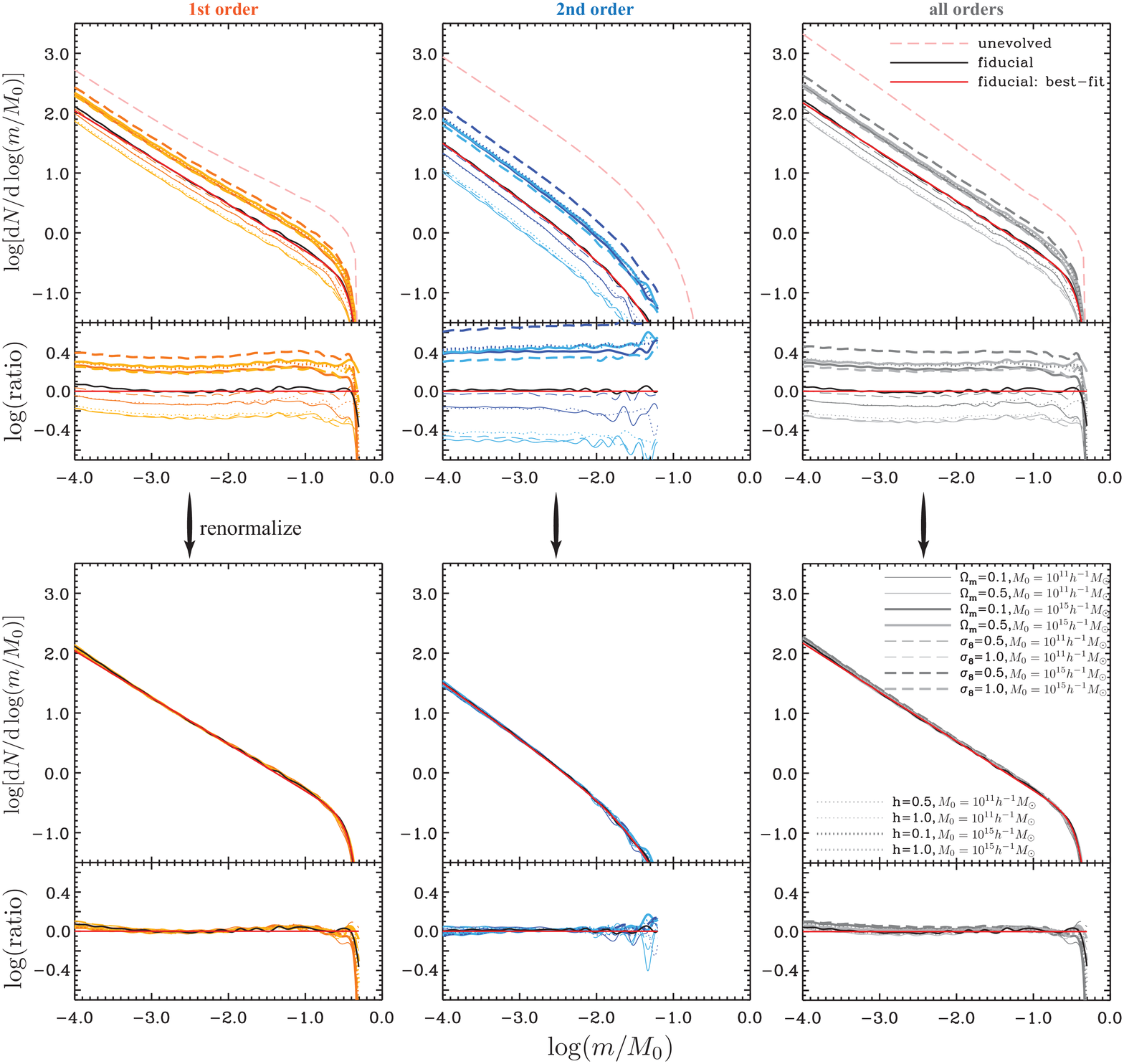,width=0.8\hdsize}}
\caption{The average SHMFs at redshift $z=0$ for different cosmologies
  and host halo masses obtained by averaging over 20,000 merger trees
  each. Panels in the upper and lower rows show the results before and
  after renormalization by the subhalo mass fraction $f_\rms$, as
  described in the text. Panels in the left, middle, and right columns
  show the results for subhaloes of first order, second order, and all
  orders. Curves of different line style and thickness correspond to
  different cosmology and halo mass, as indicated.  The solid black
  curves represent the fiducial `baseline' model, which corresponds to
  a host halo of mass $M_0 = 10^{13} \Msunh$ in the Rhapsody
  cosmology.  The dashed, light-red curves in the upper panels of the
  upper row represent the universal {\it unevolved} SHMFs. Solid, red
  curves are the fitting functions of the form of
  Eq.~(\ref{Eq:SHMFfittingFunction}) that best-fit the fiducial baseline
  model; the bottom panels in each row show the ratios with respect to
  this fitting function.}
\label{Fig:SHMFMassFracRenormalized}
\end{figure*}

The upper panels of Fig.\ref{Fig:SHMFMassFracRenormalized} plot the
SHMFs for $z=0$ host haloes with masses of $10^{11}\Msunh$ and
$10^{15}\Msunh$ for the 6 extreme cosmologies considered here. The
left, middle, and right columns show the results for first-order
subhaloes, second-order subhaloes, and subhaloes of all-orders,
respectively.  The dashed, light-red curves are the corresponding
unevolved SHMFs, which are virtually identical for all masses and
cosmologies shown. As expected, the evolved SHMFs also have very
similar shapes, but normalizations that differ by up to $\sim 1.1$dex.
Upon inspection, one can discern that the normalization of the evolved
SHMF increases with increasing $\Omega_{\rmm}$ and $M_0$, and with
decreasing $h$ and $\sigma_8$.

The normalization of the SHMF can be characterized by the total mass
fraction in subhaloes with masses $m \geq \psi_{\rm res} M_0$:
\begin{equation} \label{Eq:MassFrac}
f_\rms \equiv \int_{\psi_{\rm res}}^{1} \psi {\rmd N\over \rmd \psi}\rmd \psi 
= \int_{\psi_{\rm res}}^{1} {\rmd N\over \rmd \ln \psi}\rmd \psi,
\end{equation}
where we have used $\psi$ as shorthand for $m/M_0$. Throughout this
section we adopt a mass resolution of $\psi_{\rm res} = 10^{-4}$. In
the lower panels of Fig.~\ref{Fig:SHMFMassFracRenormalized}, we have
renormalized the SHMFs in the upper panels by multiplying $\rmd N
/\rmd\log(m/M_0)$ with the factor $f_{\rm s,fid}/f_\rms$, where
$f_{\rms,\rm{fid}}$ is the subhalo mass fraction of our fiducial
`baseline' (i.e., a host halo of mass $M_0 = 10^{13} \Msunh$ in the
Rhapsody cosmology). Clearly, this renormalization brings all SHMFs in
excellent agreement with each other; the small discrepancies at the
massive end are consistent with being due to statistical noise.
Hence, we conclude that the evolved SHMFs (of any given order)
constitute a universal, one-parameter family.

\begin{figure*}
\centerline{\psfig{figure=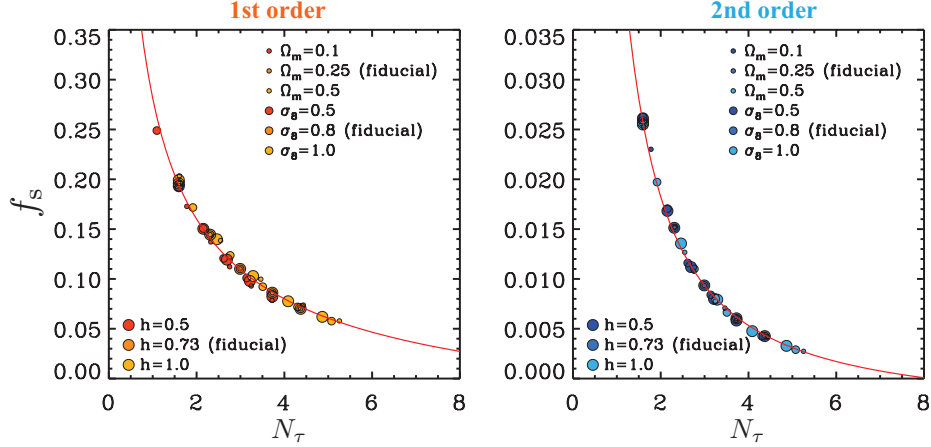,width=0.7\hdsize}}
\caption{The relation between subhalo mass fraction $f_\rms$ and the
  host halo's `dynamical age', $N_\tau$, defined as the number of
  dynamical time scales elapsed, since the formation of the (average)
  host halo (see Eq.~[\ref{Eq:Ntau}]).  The left- and right-hand
  panels correspond to first- and second-order subhaloes,
  respectively.  Filled circles of different colors and sizes
  represent different cosmologies and host halo masses, as indicated,
  while the solid, red lines are the best-fit relations of
  Eqs.~(\ref{Eq:MassFracNdynRelation1st})
  and~(\ref{Eq:MassFracNdynRelation2nd}). }
\label{Fig:SHMFcosmologyDependence}
\end{figure*}

This family of functions can be well fitted by a Schechter-like
function of the form
\begin{equation} \label{Eq:SHMFfittingFunction}
{\rmd N \over \rmd\ln\psi} = \gamma
\psi^{\alpha}\exp\left(-\beta\psi^\omega\right)\,,
\end{equation}
where $\alpha$, $\beta$, $\gamma$ and $\omega$ are free
parameters. The best-fit values for the power-law slope $\alpha$ are
$-0.78$, $-0.93$, and $-0.82$ for the SHMFs of first-order,
second-order and all-orders, respectively. Note that studies of
subhalo mass functions based on $N$-body simulations have reported
values for $\alpha$ for the SHMFs of all-orders that cover the entire
range from $-0.7$ to $-1.1$. As discussed in detail in Paper~II, this
large range owes partially to limited quality of the simulation data,
but also to the fact that different studies have used different
subhalo finders. The best-fit parameters for $\beta$ and $\omega$ are
somewhat degenerate, but we obtain good fits for $(\beta,\omega) =
(50,4)$ for SHMFs of first-order and all-orders and $(25,1)$ for
second-order subhaloes. Finally, the normalization constant $\gamma$
is related to the subhalo mass fraction, $f_\rms$, via:
\begin{equation} \label{Eq:gamma}
\gamma = { \omega\beta^{s} \over \Gamma[s,\beta\psi_{\rm res}^\omega] 
-  \Gamma[s,\beta] } \,  f_\rms ,
\end{equation}
with $\Gamma(a,x)$ the incomplete gamma function, and $s\equiv
(1+\alpha)/\omega$. The red, solid curves in
Fig.~\ref{Fig:SHMFMassFracRenormalized} represent these best-fit
functions to the fiducial `baseline' SHMF, while the smaller panels
show the ratios with respect to this fiducial fitting function. These
reveal a very weak, but systematic, upturn at the low-mass end for
SHMFs of first-order and all-orders.  This upturn is not captured by
the simple power-law behavior of Eq.~(\ref{Eq:SHMFfittingFunction}). A
similar, albeit more pronounced upturn, is also present in the {\it
  unevolved}, first-order SHMF, which, as shown by JB14, is therefore
better described by a double-Schechter-like function, $\rmd
N/\rmd\ln\psi = (\gamma_1 \psi^{\alpha_1} + \gamma_2 \psi^{\alpha_2})
\exp(-\beta\psi^\omega)$. Although the extra degrees of freedom of
such a fitting function would also improve the quality of the fits to
our inferred SHMFs, we don't believe this is warranted by the level of
accuracy of our model. We therefore choose to describe the evolved
SHMFs with the simpler Eq.~(\ref{Eq:SHMFfittingFunction}).

With the parameters $\alpha$, $\beta$ and $\omega$ specified, what
remains is to obtain an easy-to-use characterization of the
normalization parameter $\gamma$, or equivalently, of the subhalo mass
fraction, $f_\rms$. The fraction of a halo's mass that is locked up in
substructure is the outcome of a competition between halo accretion
and halo stripping. For any particular subhalo, the mass that remains
bound is basically determined by how long it has been exposed to tidal
stripping. Since the {\it unevolved} SHMF is universal, it is
therefore natural to suspect that the subhalo mass fraction, $f_\rms$,
of a host halo is closely correlated to its `age', expressed in units
of the dynamical time. For a host halo of mass $M_0$ at redshift $z_0$
this can be defined as
\begin{equation} \label{Eq:Ntau}
N_\tau(M_0,z_0) \equiv \int_{t_0}^{t(z_{\rm form})}\frac{\rmd t}{\tau_{\rm dyn}(t)}\,.
\end{equation}
Here $t$ is the lookback time, $\tau_{\rm dyn}(t)$ is the
corresponding dynamical time given by Eq.~(\ref{mytau}), $t_0$ is the
lookback time to redshift $z_0$, and $z_{\rm form} = z_{\rm
  form}(M_0,z_0)$ is the halo's formation redshift, which we define as
the redshift at which the main progenitor of the halo has reached a
mass $M_0/2$. As we demonstrate below, formulating the age of the halo
in terms of the elapsed number of dynamical times captures all the
relevant mass, redshift and cosmology dependence. 

In order to compute the formation redshift, $z_{\rm form}$, we use the
model of Giocoli \etal (2012; hereafter G12), which yields, for any cosmology, 
the median redshift $z_f$ at which the main progenitor of a halo of mass
$M_0$ at redshift $z_0$ has reached a mass $f M_0$. This requires
solving
\begin{equation} \label{Eq:zf}
\delta_\rmc(z_f) = \delta_\rmc(z_0) + \tilde{w}_f \sqrt{\sigma^2(fM_0)-\sigma^2(M_0)},
\end{equation}
where $\tilde{w}_f = \sqrt{2\ln(\alpha_f + 1)}$ and $\alpha_f =
0.815e^{-2f^3}/f^{0.707}$. Here $\sigma^2(M)$ is the mass variance and
$\delta_\rmc(z) \equiv 1.686/D(z)$ with $D(z)$ the linear growth rate
normalized to unity at $z=0$. We define $z_{\rm form}$ as the solution
of Eq.~(\ref{Eq:zf}) for $z_f$ with $f=0.5$.

Fig.~\ref{Fig:SHMFcosmologyDependence} plots the subhalo mass
fraction, $f_\rms$, obtained from the evolved SHMFs shown in
Fig.~\ref{Fig:SHMFMassFracRenormalized}, as function of the halo `age'
$N_{\tau}$, computed using Eq.~(\ref{Eq:Ntau}) with $z_{\rm form}$
obtained from the G12 model as described above. As expected, these two
quantities are tightly correlated.  We find that the subhalo mass
fractions for first and second order subhaloes are accurately
described by
\begin{equation} \label{Eq:MassFracNdynRelation1st}
f_{\rm s,1st} = {0.3563 \over N_\tau^{0.6}} - 0.075\,,
\end{equation}
and
\begin{equation} \label{Eq:MassFracNdynRelation2nd}
f_{\rm s,2nd} = {0.0535 \over  N_\tau^{1.3}} - 0.0035\,,
\end{equation}
respectively, which are indicated by the red lines. We emphasize that
these relations are valid for any halo mass, $M_0$, any redshift,
$z_0$, and any reasonable $\Lambda$CDM cosmology. Note that, since our
halo mass definition is `inclusive', the first-order mass fraction
$f_{\rm s,1st}$ also describes the mass fraction in subhaloes of all
orders.  Also, recall that the mass fractions in
Eqs.~(\ref{Eq:MassFracNdynRelation1st})
and~(\ref{Eq:MassFracNdynRelation2nd}) correspond to a mass resolution
of $\psi_{\rm res} = 10^{-4}$. This needs to be taken into account
when using these mass fractions to compute the SHMF normalization
parameter, $\gamma$, with the help of Eq.~(\ref{Eq:gamma}).

The above relations between $f_\rms$ and $N_\tau$ allow one to write
down the average, evolved SHMF for a halo of arbitrary mass, at any
redshift, and for a wide range of $\Lambda$CDM cosmologies. We now
examine whether similar, universal fitting functions can be found to
describe both the evolved and unevolved SHVFs.

\subsection{The Unevolved SHVF} 
\label{Sec:UnevolvedSHVF}

\begin{figure*}
\centerline{\psfig{figure=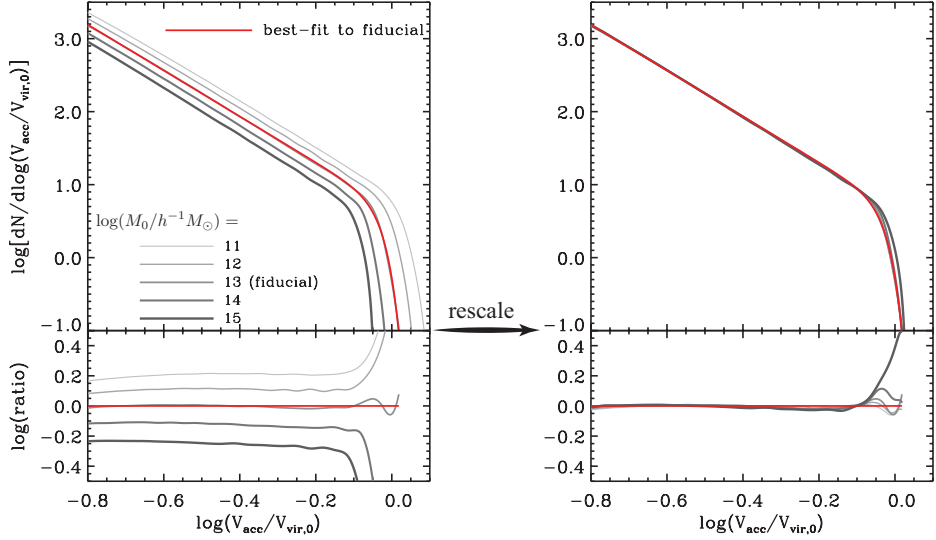,width=0.7\hdsize}}
\caption{Panels on the left show the unevolved subhalo velocity
  functions, $\rmd N/\rmd\log(V_{\rm acc}/V_{\rm vir,0})$, for five
  different host halo masses, as indicated, in the Rhapsody
  cosmology. Each of these has been obtained by averaging 20,000
  merger trees. In the panels on the right, these have been rescaled
  using the scale parameter $a$ given by Eq.~(\ref{Eq:a}). In both
  panels the red curves correspond to the fitting function of the form
  of Eq.~(\ref{Eq:SHVFfittingFunction}) that best-fit the results for
  $M_0 = 10^{13} \Msunh$, which has best-fit parameters $\alpha =
  -3.2$, $\beta = 2.2$, $\gamma = 2.05$, $\omega=13$, and (by
  definition) $a=1$. The bottom panel plots the ratios relative to
  this fitting function.}
\label{Fig:USVFmassDependence}
\end{figure*}

The unevolved SHVF, ${\rm d}N/{\rm d}\ln(V_{\rm acc}/V_{\rm vir,0})$,
provides the basis for the popular technique of subhalo abundance
matching.  Unfortunately, the convenient universality present in the
case of the unevolved mass function does not hold for the unevolved
velocity function. This is easy to understand from the fact that the
relation between $m_{\rm acc}$ and $V_{\rm acc} = V_{\rm max}(z_{\rm
  acc})$ depends on the concentration-mass-redshift relation. Hence,
the `mapping' of $m_{\rm acc}/M_0$ into $V_{\rm acc}/V_{\rm vir,0}$
depends on mass, redshift and cosmology.

The left-hand panels of Fig.~\ref{Fig:USVFmassDependence} show the
unevolved SHVFs of all-orders for host haloes of different masses in
the Rhapsody cosmology, as indicated.  These have been obtained
averaging over 20,000 Monte-Carlo realizations. Although both the
normalization and the scale at which the transition to exponential
decay occurs differ from one SHVF to the other, the shapes still look
close to universal. Hence, it is natural to describe the unevolved
velocity function with a functional form 
\begin{equation} \label{Eq:SHVFfittingFunction}
{\rmd N \over \rmd \ln\psi} = \gamma \left(a\psi\right)^\alpha 
\exp\left[-\beta\left(a\psi\right)^\omega\right],
\end{equation}
where we have introduced a scale parameter, $a$, in addition to the
free parameters $\alpha$, $\beta$, $\gamma$, and $\omega$. We now seek
to find a parameterization of $a$ for which the unevolved SHVF is
(close to) universal, i.e., for which the other parameters are
independent of host mass, redshift and cosmology.

To do so, we use the fact that 
\begin{equation} \label{Eq:link}
{V_{\rm max}(m_{\rm acc},z_{\rm acc}) \over V_{\rm vir}(M_0,z_0)} = 
{V_{\rm max}(m_{\rm acc},z_{\rm acc}) \over V_{\rm vir}(m_{\rm acc},z_0)}
\times
{V_{\rm vir}(m_{\rm acc},z_0) \over V_{\rm vir}(M_0,z_0)}\,.
\end{equation}
Since the second factor on the right-hand side only depends on $m_{\rm
  acc}/M_0$, and has no dependence on cosmology or redshift, and since
the mass function of $m_{\rm acc}/M_0$ is universal, a logical choice
for the scale factor $a$ is $V_{\rm vir}(\langle m_{\rm
  acc}\rangle,z_0) / V_{\rm max}(\langle m_{\rm acc} \rangle,\langle
z_{\rm acc} \rangle)$ where $\langle m_{\rm acc} \rangle$ and $\langle
z_{\rm acc} \rangle$ are representative values for the mass and
redshift at accretion.  For the latter we use the redshift $z_f$ by
which the main progenitor of the host halo has assembled a faction $f$
of its final mass $M_0$. For the characteristic subhalo mass at
accretion we take $\langle m_{\rm acc} \rangle = M_0/10$, which is
close to the exponential cut-off scale of the unevolved SHMF. Hence, we
have that
\begin{equation} \label{Eq:a}
a = C \, {V_{\rm vir}(0.1fM_0,z_0) \over V_{\rm max}(0.1fM_0,z_f)},
\end{equation}
where we have introduced the normalization $C$ which we tune such that
$a=1$ for our fiducial `baseline' model. For a given value of $f$, one
can use the G12 model to compute the corresponding $z_f$, and use
Eqs.~(\ref{Vvir}) and (\ref{VmaxHost}) to compute $V_{\rm
  vir}(0.1fM_0,z_0)$ and $V_{\rm max}(0.1fM_0,z_f)$ , respectively.
The latter requires the halo concentration parameter, $c$, which we
compute using the concentration-mass-redshift relation of Zhao \etal
(2009).

We have experimented with different values for $f$, and obtain the
best results for $f=0.25$, for which $C=1.536$. The right-hand panels
of Fig.~\ref{Fig:USVFmassDependence} plot $\rmd N/\rmd\log(V_{\rm
  acc}/V_{\rm vir,0})$ versus $a\,V_{\rm acc}/V_{\rm vir,0}$, where
$a$ is computed using Eq.~(\ref{Eq:a}) with these best-fit values.  It
is clear that this simple rescaling accounts for virtually all the
halo mass dependence in the unevolved SHVF.  The rescaled, unevolved
SHVFs are almost indistinguishable from each other and are well fit by
Eq.~(\ref{Eq:SHVFfittingFunction}) with $\alpha=-3.2$, $\beta=2.2$,
$\gamma=2.05$, $\omega=13$, and $a=1$ (red curve). Some discrepancies
that go beyond statistical noise are evident at the high-velocity end
($aV_{\rm acc}/V_{\rm vir,0} \gta 0.9$), but only for the most massive
host haloes. Although the discrepancies are within a factor of two, we
caution that the `universal' unevolved SHMF presented here is less
reliable for the most massive subhaloes in the most massive
(cluster-sized) host haloes.
 
We have also verified that the same scale parameter $a$ (with
$f=0.25$) also describes the cosmology- and redshift- dependences
quite well. There is a weak dependence of the power-law slope $\alpha$
on cosmology, but this only becomes cumbersome when the cosmological
parameters deviate substantially (i.e., by about a factor of two or
more) from the `baseline' Rhapsody cosmology with $(\Omega_{\rmm},
h, \sigma_8) = (0.25, 0.73, 0.8)$. Overall, though, the rescaling
introduced here, and summarized in Appendix~A, allows one to compute
reliable, unevolved subhalo velocity functions for host haloes of
different mass, at different redshifts, and for a broad range of
$\Lambda$CDM cosmologies.

\subsection{The Evolved SHVF} 
\label{Sec:EvolvedSHVF}

\begin{figure*}
\centerline{\psfig{figure=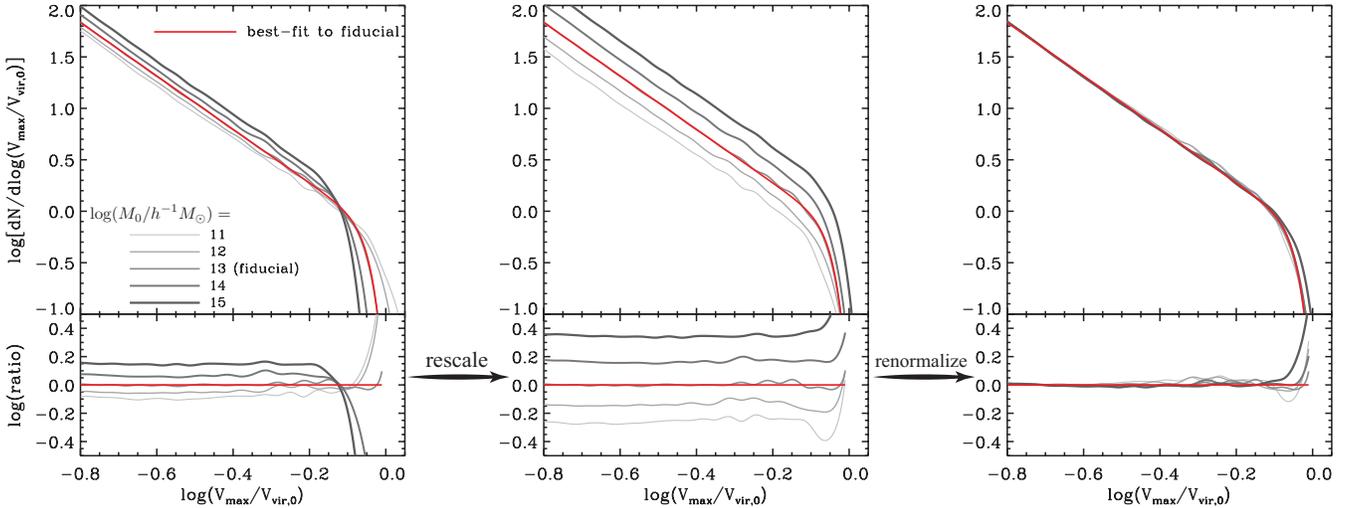,width=1.0\hdsize}}
\caption{Same as Fig.~\ref{Fig:USVFmassDependence}, but here we show
  the corresponding results for the {\it evolved} SHVFs, $\rmd
  N/\rmd\log(V_{\rm max}/V_{\rm vir,0})$. After rescaling with the
  scale parameter $a$, the various SHVFs have a universal shape, but
  different normalizations (panels in middle column). This dependence
  can be renormalized by scaling the SHVFs with a factor $f_\rms^{1.4}$
  (panels in right-hand column). After this two-step process of
  rescaling and renormalization, the evolved SHVFs are accurately
  described by Eq.~(\ref{Eq:SHVFfittingFunction}) with $\alpha =
  -2.6$, $\beta=4$, $\gamma = 0.248$, and $\omega=15$ (red
  curves).}
\label{Fig:ESVFmassDependence}
\end{figure*}

The left-hand panels of Fig.~\ref{Fig:ESVFmassDependence} plot the
{\it evolved} SHVFs corresponding to the {\it unevolved} SHVFs shown
in Fig.~\ref{Fig:USVFmassDependence}. Similar to the evolved {\it
  mass} functions, the power-law slopes of the evolved {\it velocity}
functions are independent of host halo mass, and the normalization
increases with increasing $M_0$. However, at the high-velocity end
this halo mass-dependence flips over, and the exponential cut-off
occurs at smaller $V_{\rm max}/V_{\rm vir,0}$ for more massive host
haloes. As we demonstrate in Paper~II, these trends are consistent
with results from $N$-body simulations.

In an attempt to construct a universal fitting function for the {\it
  evolved} SHVF, we first apply the rescaling that successfully
describes the non-universality of the {\it unevolved} SHVF.  This
results in the rescaled velocity functions shown in the middle panels
of Fig.~\ref{Fig:ESVFmassDependence}. The cut-off now occurs at
roughly the same scale, which has removed the `cross-over' present in
the left-hand panels. The normalizations, though, still depend on host
halo mass, which reflects the impact of subhalo mass evolution. Hence,
we expect that this normalization scales with the halo's dynamical
age, $N_\tau$. Recall that in the case of the evolved mass function,
the normalization constant is simply proportional to the subhalo mass
fraction $f_\rms$, which, after all, is simply an integral of the mass
function. In the case of the evolved SHVF, we assume that the
normalization scales with $f_\rms^b$, where $b$ is some constant. By
trial and error we find that $b=1.4$ accurately captures the scaling of
the SHVFs. This is demonstrated in the right-hand panels of
Fig.~\ref{Fig:ESVFmassDependence}, which indicates that
renormalization by $f_\rms^{1.4}$ yields SHVFs that are almost
indistinguishable. This universal, evolved SHVF is well described by
Eq.~(\ref{Eq:SHVFfittingFunction}) with $(\alpha, \beta, \gamma,
\omega) = (-2.6, 4, 0.248, 15)$.  As with the {\it unevolved}
SHVF, small discrepancies are evident at the high-velocity end for the
most massive host haloes, but overall this two-stage process of
rescaling and renormalization nicely captures the mass-, redshift- and
cosmology dependence of the evolved SHVF for subhaloes of all-orders
(see also Appendix~A).

%


\section{Discussion} \label{Sec:Discussion}

The results presented in this paper show that the orbit-averaged
subhalo mass-loss rates are accurately described by
\begin{equation}\label{mydecay_repeat}
\dot{m} = - \calA \, {m \over \tau_{\rm dyn}} \left({m\over M}\right)^{\zeta}.
\end{equation}
with $\tau_{\rm dyn}$ the halo's (instantaneous) dynamical time,
$\zeta = 0.07$, and $\calA$ a random variable that follows a
log-normal distribution with median $\bar{\calA} = 1.34$ and
dispersion $\sigma_{\log\calA} = 0.17$. This implies that, in an
orbit-averaged sense, dark matter subhaloes evolve as
\begin{equation}\label{mt}
m(t) = m_{\rm acc} \, \left[1 + \zeta \, \calA \, \left({m_a \over M_0}\right)^{\zeta}
\, \left\{\tilde{N}_{\tau}(t_{\rm acc}) - \tilde{N}_{\tau}(t)\right\} \right]^{-1/\zeta}
\end{equation}
where $t$ is {\it lookback time}, $t_{\rm acc}$ and $m_{\rm acc}$ are the
lookback time and subhalo mass at accretion, $M_0$ is the present day
mass of the host halo, and
\begin{equation}\label{tildeNtau}
\tilde{N}_\tau(t) \equiv \int_0^t \left[{M(t) \over M_0}\right]^{-\zeta} \,
{\rmd t \over \tau_{\rm dyn}(t)}
\end{equation}
is some measure for the number of dynamical times that have elapsed in
an evolving dark matter halo since lookback time $t$.

It is interesting to see what this implies for the amount of mass that
is stripped from a typical subhalo during its first radial orbit. In
units of the mass of the subhalo at infall, this is given by
\begin{equation}\label{Mstrip}
{\Delta m \over m_{\rm acc}} \equiv {{m_{\rm acc} - m(t_{\rm acc} - T_\rmr)} \over m_{\rm acc}} 
\end{equation}
with $T_\rmr$ the radial orbital period given by Eq.~(\ref{Tr}).
Without loosing generality, we set $t_{\rm acc} = T_\rmr$, so that the
subhalo has just completed its first radial orbit at the present
day. In this case, we have that
\begin{equation}\label{Mstripdet}
{\Delta m \over m_{\rm acc}} =  1 -  
\left[1 + \zeta \calA \left({m_a \over M_0}\right)^{\zeta}
\, \tilde{N}_{\tau}(T_\rmr)\right]^{-1/\zeta}
\end{equation}
Using the toy model described in \S\ref{Sec:ToyModel} we find that the
distribution of $T_\rmr$ at $z=0$ is close to uniform over the
interval $[5,9]$ Gyr, which has its origin in the uniform distribution
of $R_\rms$ ($T_\rmr$ depends strongly on $E$ but has very little
dependence on $L$).

\begin{figure*}
\centerline{\psfig{figure=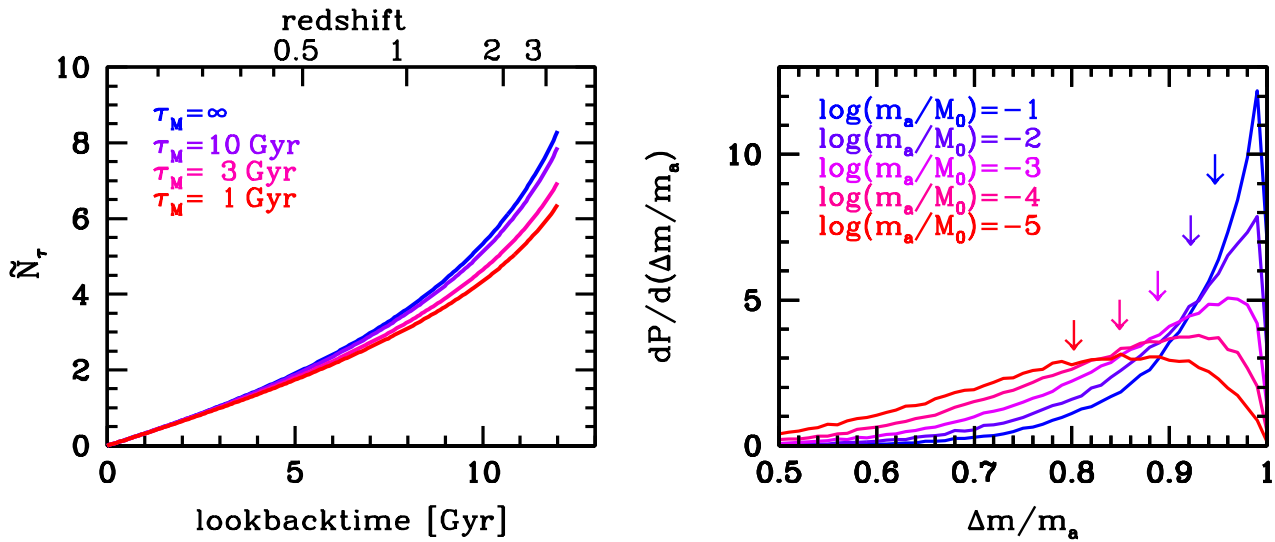,width=0.95\hdsize}}
\caption{{\it Left-hand panel:} the quantity $\tilde{N}_\tau$, defined
  by Eq.~(\ref{tildeNtau}), as function of lookback time $t$ for four
  different values of the time-scale for halo mass growth, $\tau_M$,
  as indicated. As is apparent, $\tilde{N}_\tau$ is not very sensitive
  to how host halo masses grow over time. This is a manifestation of
  the small value for $\zeta$, which indicates that subhalo mass loss
  rates depend only weakly on host halo mass. {\it Right-hand panel:}
  Distributions of the fractional subhalo mass lost during the first
  radial orbit after infall, $\Delta m/m_{\rm acc}$. Results are shown for
  five different values of $m_{\rm acc}/M_0$, as indicated. Arrows indicate
  the medians of the corresponding distributions. Note that subhaloes,
  on average, lose more than 80 percent of their infall mass during
  their first radial orbit.  All these results are for a Planck
  cosmology with $(\Omega_{\rmm},h) = (0.318,0.671)$, but results are
  very similar for other $\Lambda$CDM cosmologies that are consistent
  with current observational constraints.}
\label{Fig:Ntau}
\end{figure*}

The left-hand panel of Fig.~\ref{Fig:Ntau} plots $\tilde{N}_{\tau}$ as
function of lookback time, $t$, where we have assumed, for simplicity,
that dark matter haloes grow in mass exponentially on a time scale
$\tau_M$, i.e., $M(t) = M_0 \, \exp(-t/\tau_M)$. We also assumed a
`Planck cosmology' with $\Omega_\rmm = 0.318$, $\Omega_{\Lambda} =
0.682$ and $h=0.671$, but we emphasize that the results are almost
indistinguishable for other, similar cosmologies, such as those
advocated by different data releases of the WMAP experiment.  Results
are shown for four different values of $\tau_M$, ranging from infinity
(i.e., no evolution in host halo mass) to $\tau_M = 1$Gyr (i.e., host
halo mass has grown by almost a factor three during the last Gyr).
This more than covers the range of growth rates of dark matter haloes
in the mass range $10^{11} \Msunh < M_0 < 10^{15}\Msunh$. As is
evident, the interval $T_\rmr \in [5,9]$Gyr translates roughly into
$\tilde{N}_\tau(T_\rmr) \in [2,4]$, with very little dependence on
$\tau_M$\footnote{This also implies that the results presented here
  are insensitive to deviations of $M(t)/M_0$ from an exponential}.
Hence, the typical radial orbital period of a subhalo following infall
lasts roughly 2 to 4 dynamical times. This may sound somewhat
counter-intuitive, but note that the dynamical time is an average for
the entire halo, which is not representative of orbits at first infall.

The right-hand panel of Fig.~\ref{Fig:Ntau} plots the distribution of
$\Delta m/m_{\rm acc}$ for the same Planck cosmology, obtained using
Eq.~(\ref{Mstripdet}) with $\zeta = 0.07$ and $\tau_M = 10$Gyr (roughly
representative for a Milky-Way sized dark matter halo, though the
results only depend very weakly on $\tau_M$). The orbital periods,
$T_\rmr$, are sampled from a uniform distribution covering the range
from 5 to 9 Gyr, while the mass-loss rate normalization parameter,
$\calA$, is sampled from the log-normal given by Eq.~(\ref{ProbA})
with $\bar{\calA} = 1.34$.  Results are shown for five different
values of $m_{\rm acc}/M_0$, as indicated. The medians of the distributions
are indicated by arrows, and range from $0.80$ for $m_{\rm acc}/M_0 =
10^{-5}$ to $0.95$ for $m_{\rm acc}/M_0 = 10^{-1}$. Sampling $m_{\rm acc}/M_0$
from the actual unevolved SHMF for $m_{\rm acc}/M_0 \geq 10^{-5}$ yields a
distribution that is intermediate between those for $m_{\rm acc}/M_0 =
10^{-4}$ and $m_{\rm acc}/M_0 = 10^{-5}$ with a median of $0.827$.  Note
that only a minute fraction of subhaloes is expected to hang on to
more than 50 percent of their infall mass after one radial orbit. Hence,
{\it subhaloes lose the vast majority (typically more than 80
  percent) of their mass during their very first radial orbit}. We
emphasize that most of this mass loss is likely to occur near
pericenter (and hence, roughly a time $T_\rmr/2$ after infall), but we
caution that our model only treats orbit-averaged mass-loss rates, and
should therefore not be used to make predictions regarding mass-loss
rates on significantly shorter time-scales.

The dependence of $\Delta m/m_{\rm acc}$ on $m_{\rm acc}/M_0$ owes to two
effects: (i) the concentration-mass relation of dark matter haloes,
which makes subhaloes with a lower value of $m_{\rm acc}/M_0$ relatively
denser compared to its host halo, and therefore more resilient to
tidal stripping, and (ii) dynamical friction, which will cause more
massive subhaloes to lose more orbital energy and angular momentum,
reducing their pericentric distance, and thus causing enhanced
stripping. However, with the dramatic mass stripping rates revealed
here, it is also clear that dynamical friction cannot play a very
important role after first pericentric passage; as a rule of thumb,
the dynamical friction time is only shorter than the Hubble time if
$m/M \gta 0.1$ (e.g., Mo, van den Bosch \& White 2010).  Even if a
subhalo is that massive at infall, it is very likely to be stripped
below this limit after its first pericentric passage. Hence, mass
stripping is a far more important process for the evolution of dark
matter subhaloes than dynamical friction (see also Taffoni \etal 2003;
Taylor \& Babul 2004; Pe\~narrubia \& Benson 2005; Zentner \etal 2005;
Gan \etal 2010), and one does not make large errors by ignoring
dynamical friction altogether.


\section{Summary} 
\label{Sec:Summary}

We have presented a new semi-analytical model that uses EPS merger trees to
generate evolved subhalo populations. The model is based on the method
pioneered by B05, and evolves the mass of dark matter haloes using a
simple model for the {\it orbit-averaged} subhalo mass loss rate. This
avoids having to integrate individual subhalo orbits, as done in other
semi-analytical models for dark matter substructure (e.g., Taylor \&
Babul 2004, 2005a,b; Benson \etal 2002; Taffoni \etal 2003; Oguri \&
Lee 2004; Zentner \& Bullock 2003; Pe\~{n}arrubia \& Benson 2005;
Zentner \etal 2005; Gan \etal 2010). We have made a number of
improvements and extensions with respect to the original B05 model; in
particular, we
\begin{enumerate}

\item use Monte Carlo merger trees constructed using the method of
  P08, which, as demonstrated in JB14, yields
  results in much better agreement with numerical simulations than the
  Somerville \& Kolatt (1999) method used by B05.

\item construct and use complete merger trees, rather than just the
  mass assembly histories of the main progenitor. This allows us to
  investigate the statistics of subhaloes of different orders.

\item adopt a new mass loss model, that is calibrated against
  numerical simulations and which also accounts for the scatter in
  subhalo mass loss rates that arises from scatter in orbital
  properties (energy and angular momentum) and (sub)halo
  concentrations.

\item include a method for converting halo mass to maximum circular
  velocity, thus allowing us to study subhalo velocity functions as
  well as subhalo mass functions.

\end{enumerate}

In this paper, the first in a series that addresses the statistics of
dark matter subhaloes, we have mainly focussed on the {\it average}
subhalo mass and velocity functions, where the average is taken over
large numbers of Monte Carlo realizations for a certain host halo
mass, $M_0$, redshift, $z_0$, and cosmology. Our model has only one
free parameters, which sets the overall normalization of the
orbit-averaged mass loss rates of dark matter subhaloes. After tuning
this parameter such that the model reproduces the normalization of the
evolved SHMF in the numerical simulations of G08, the
same model can accurately reproduce the evolved subhalo mass and
velocity functions in numerical simulations for host haloes of
different mass, in different $\Lambda$CDM cosmologies, and for
subhaloes of different orders, without having to adjust this
parameter.

The inferred orbit-averaged mass loss rates are consistent with the
simulation results of G08, and imply that an average dark matter
subhalo loses in excess of 80 percent of its infall mass during its
first radial orbit within the host halo. More massive subhaloes, in
units of the normalized mass, $m/M$, lose their mass more rapidly due
to (i) the concentration-mass relation of dark matter haloes, which
causes subhaloes with smaller $m/M$ to be more resilient to tidal
stripping, and (ii) dynamical friction, which causes more massive
subhaloes to lose more orbital energy and angular momentum, resulting
in enhanced stripping. According to our mass loss model, subhaloes
with an infall mass that is 10 percent of the host halo mass will
lose on average more than 95 percent of their infall mass during
their first radial orbital period.

One of the main findings of this paper is that the average subhalo
mass and velocity functions, both evolved and unevolved, can be
accurately fit by a simple Schechter-like function of the form
\begin{equation} \label{fitgeneral}
{\rmd N \over \rmd \ln \psi} = \gamma \, (\psi)^{\alpha} \,
\exp\left[-\beta(\psi)^\omega\right]\,.
\end{equation}
where, depending on which function is being considered, $\psi$ is
$m/M_0$, $m_\rma/M_0$, $V_{\rm max}/V_{\rm vir,0}$, or $V_{\rm
  acc}/V_{\rm vir,0}$. In particular, restricting ourselves to
$\Lambda$CDM cosmologies with parameters that are consistent with
recent constraints within a factor of roughly two, we find that
\begin{itemize}

\item The {\it unevolved} SHMF is (close to) universal, with the
  parameters $(\alpha, \beta, \gamma, \omega)$ independent of host
  halo mass, redshift and cosmology (see also B05; Li \& Mo 2009; Yang
  \etal 2011). We emphasize, though, that although the functional form
  of Eq.~(\ref{fitgeneral}) can adequately describe this univeral
  unevolved SHMF, it is more accurately described by the
  double-Schechter-like function presented in JB14.

\item The {\it evolved} SHMF has a universal shape (i.e., fixed
  $\alpha$, $\beta$ and $\omega$), which is accurately described by
  Eq.~(\ref{fitgeneral}), but with a normalization, $\gamma$, that
  depends on host halo mass, redshift and cosmology. We have
  demonstrated that $\gamma$ is tightly correlated with the `dynamical
  age' of the host halo, defined as the number of halo dynamical times
  that have elapsed since its formation (i.e., since redshift
  $z_{1/2}$ at which the host halo's main progenitor reaches a mass
  equal to $M_0/2$). Using this relation we have presented a universal
  fitting function for the average, evolved SHMF that is valid for any
  host halo mass, at any redshift, and for any $\Lambda$CDM cosmology.
  The corresponding power-law slopes, $\alpha$, are $-0.78$, $-0.93$
  and $-0.82$ for first-order subhaloes, second-order subhaloes (i.e.,
  sub-subhaloes), and for subhaloes of all orders, significantly
  shallower than what has been claimed in numerous studies based on
  numerical simulations (see Paper~II for a detailed discussion).

\item Unlike the unevolved mass function, the {\it unevolved} SHVF is
  not universal, in that the parameter $\beta$ is found to depend on
  host mass, redshift and cosmology. This has its origin in the
  concentration-mass-redshift relation of dark matter haloes, and can
  be accounted for by replacing $\psi$ in Eq.~(\ref{fitgeneral}) with
  $a\psi$, where $a$ is a (universal) scale factor given by $a \propto
  V_{\rm vir}(M_0/40,z_0)/V_{\rm max}(M_0/40,z_{0.25})$. When using
  this simple rescaling, one obtains a universal fitting function for
  the unevolved SHVF whose parameters $\alpha$, $\beta$, $\gamma$ and
  $\omega$ are independent of host mass, redshift and cosmology. Note
  that this unevolved SHVF is one of the key ingredients in the popular
  method of subhalo abundance matching.

\item Taking into account both the `dynamical age'-dependence of the
  normalization of the evolved SHMF and the `$a$'-scaling of the
  unevolved SHVF, also yields a universal fitting function for the
  {\it evolved} SHVF. In this case we find that the power-law slope for
  the evolved SHVF of all orders is equal to $\alpha = -2.6$. 

\end{itemize}

The various, universal fitting functions for the subhalo mass and
velocity functions presented here, and summarized in Appendix~A, can
be used to quickly compute the average abundance of subhaloes of given
mass or maximum circular velocity, at any redshift, and for any
(reasonable) $\Lambda$CDM cosmology, without having to run and analyze
high resolution numerical simulations.  In the second paper in this
series (van den Bosch \& Jiang 2014), we compare subhalo mass and
velocity functions obtained from different simulations and with
different subhalo finders, among each other, and with predictions from
our semi-analytical model. We demonstrate that our model is in excellent
agreement with simulation results that analyze their data with halo
finders that use the full 6D phase-space information (e.g., {\tt
  ROCKSTAR}), or that use temporal information (e.g., {\tt SURV}).
Results obtained using subhalo finders that only rely on the densities
in configuration space are shown to dramatically underpredict the
abundance of massive subhaloes, by more than an order of magnitude. In
the third paper in this series (Jiang \& van den Bosch, in
preparation), we use our model to investigate, in unprecedented
detail, the halo-to-halo variance of dark matter substructure, which
is important, among others, for assessing the severity of the
`too-big-to-fail' problem (see also Purcell \& Zentner 2012).


\section*{Acknowledgments}

We are grateful to Andrew Wetzel and Mike Boylan-Kolchin for helpful
discussions and to Hao-Yu Wu for sharing some of their data from the
Rhapsody simulation project. We also thank the people behind the
MultiDark simulation for making their halo catalogs publicly
available, and the organizers of the program "First Galaxies and Faint
Dwarfs: Clues to the Small Scale Structure of Cold Dark Matter" held
at the Kavli Institute for Theoretical Physics (KITP) in Santa
Barbara, for creating a stimulating, interactive environment that
started the research presented in this paper, and the KITP staff for
all their help and hospitality. This research was supported in part by
the National Science Foundation under Grant No. PHY11-25915.



\appendix

\section{Universal Fitting Functions for the Subhalo Mass and Velocity Functions}
\label{App:universal}

This appendix describes how to compute an average mass or velocity
function (evolved or unevolved) for subhaloes of all orders.  We have
shown in \S\ref{Sec:Universal} that the average evolved and unevolved
subhalo mass and velocity functions of a host halo of mass $M_0$ at
redshift $z_0$ are well described by the general fitting function
\begin{equation}\label{AppEq:GeneralFittingFunction}
{\rmd N \over \rmd\ln\psi} = \gamma \left(a\psi\right)^{\alpha} 
\exp\left[-\beta \left(a\psi\right)^\omega\right],
\end{equation}
where $\psi$ stands for the corresponding quantity; $m_{\rm acc}/M_0$,
$m/M_0$, $V_{\rm acc}/V_{\rm vir,0}$, or $V_{\rm max}/V_{\rm
  vir,0}$. For subhaloes of all orders, the corresponding best-fit
values for $\alpha$, $\beta$, $\gamma$, $\omega$ and $a$ are listed in
Table~A1. As is evident, they are completely described by the subhalo
mass fraction $f_\rms$ and the scale parameter $a$. In what follows we
outline the steps required to compute these two values for a host halo
of mass $M_0$ at redshift $z_0$ in a given $\Lambda$CDM cosmology:
\begin{enumerate}

\item Obtain the redshift $z_f$, by which the main progenitor has
  assembled a fraction $f$ of its final mass $M_0$ at redshift $z_0$.
  In particular, compute $z_f$ for $f=0.5$, $0.25$, and $0.04$ using
  the G12 model for formation redshift, by solving for the root of
\begin{equation} \label{Eq:zftwo}
\delta_\rmc(z_f) = \delta_\rmc(z_0) + \tilde{w}_f \sqrt{\sigma^2(fM_0)-\sigma^2(M_0)},
\end{equation}
  where $\tilde{w}_f = \sqrt{2\ln(\alpha_f + 1)}$ and $\alpha_f =
  0.815e^{-2f^3}/f^{0.707}$.

\item Compute the `dynamical age', $N_\tau$, of the host halo
  using Eq.~(\ref{Eq:Ntau}) with $z_{\rm form} = z_{0.5}$. Use this to
  infer the subhalo mass fraction, $f_\rms$, from
  Eq.(\ref{Eq:MassFracNdynRelation1st}).

\item Compute the scale parameter, $a$, given by
\begin{equation} \label{AppEq:a}
a = 1.536 \, {V_{\rm vir}(M_0/40,z_0) \over V_{\rm max}(M_0/40,z_{0.25})},
\end{equation}
  with $V_{\rm vir}$ and $V_{\rm max}$ given by Eqs.~(\ref{Vvir})
  and~(\ref{VmaxHost}), respectively, and with $z_{0.25}$ obtained in
  step (i). Note that the computation of $V_{\rm max}$ requires the
  halo concentration parameter, $c$, which can be computed from the
  $z_{0.04}$ obtained under step (i) using the
  concentration-mass-redshift relation of Zhao \etal (2009), given by
  Eq.~(\ref{cvir}).

\end{enumerate}

Although we believe this `recipe' to be reliable for host haloes that
cover a wide range in halo masses and redshifts, and for a wide range
of cosmologies, we caution that we have only been able to test it
against numerical simulations over the mass range $10^{11} \Msunh \lta
M_0 \lta 10^{15}\Msunh$ at relatively low redshifts, and for a range
of cosmologies that are all similar to the best-fit cosmologies
advocated by the recent CMB experiments (see Paper~II for details). We
therefore caution against using this method blindly for cosmologies
and/or halo masses that are very different from those mentioned above.
Finally, we mention that the fitting function for the {\it unevolved}
SHMF described by the parameters in Table \ref{Tab:SHMFandSHVF} is
less accurate than that presented in JB14. Hence, if accuracy is a
concern, we recommend the latter over the one presented here.
\begin{table}
\caption{Parameters for universal SHMF and SHVF.} 
\label{Tab:SHMFandSHVF}
\begin{tabular}{lccccc}
\hline
$\psi$ & $\gamma$ & $\alpha$  & $\beta$ & $\omega$ & $a$  \\ 
(1) & (2) & (3) & (4) & (5) & (6)\\
\hline \hline
$m_{\rm acc}/M_0$          & 0.22             & -0.91  &   6  & 3 & 1 \\
$m/M_0$                  & $0.31f_\rms$       & -0.82  &  50 &  4 & 1 \\
$V_{\rm acc}/V_{\rm vir,0}$ & 2.05             & -3.2  & 2.2 & 13 & Eq.(\ref{AppEq:a})\\
$V_{\rm max}/V_{\rm vir,0}$ & $5.45f_\rms^{1.4}$ & -2.6 & 4 & 15 & Eq.(\ref{AppEq:a})\\
\hline
\end{tabular}
\medskip
\end{table}

\label{lastpage}


\end{document}